\documentclass[fleqn,12pt]{wlscirep}
\usepackage{graphicx}
\usepackage{amssymb}
\usepackage{newfloat}
\usepackage{textcomp}
\DeclareFloatingEnvironment[name={Extended Data Figure}]{extfig}
\newcommand{\mbh}{$M_\bullet$}
\newcommand{\msun}{\hbox{M$_{\odot}$}}

\title{Anisotropic satellite galaxy quenching modulated by supermassive black hole activity}

\author[1,2,3,*]{Ignacio Mart\'in-Navarro}
\author[3]{Annalisa Pillepich}
\author[4,5]{Dylan Nelson}
\author[6]{Vicente Rodriguez-Gomez}
\author[3]{Martina Donnari}
\author[7]{Lars Hernquist}
\author[4]{Volker Springel}

\affil[1]{Instituto de Astrof\'isica de Canarias, V\'ia L\'actea s/n, E-38205 La Laguna, Tenerife, Spain}
\affil[2]{Departamento de Astrof\'isica, Universidad de La Laguna, E-38205 La Laguna, Tenerife, Spain}
\affil[3]{Max-Planck Institut f\"ur Astronomie, Konigstuhl 17, D-69117 Heidelberg, Germany}
\affil[4]{Max-Planck-Institut f\"{u}r Astrophysik, Karl-Schwarzschild-Str. 1, D-85748 Garching, Germany}
\affil[5]{Universit\"{a}t Heidelberg, Zentrum f\"{u}r Astronomie, Institut f\"{u}r theoretische Astrophysik, Albert-Ueberle-Str. 2, 69120 Heidelberg, Germany}
\affil[6]{Instituto de Radioastronom{\'i}a y Astrof{\'i}sica, Universidad Nacional Aut{\'o}noma de M{\'e}xico, 58089 Morelia, M{\'e}xico}
\affil[7]{Harvard-Smithsonian Center for Astrophysics, 60 Garden Street, Cambridge, MA, 02138, USA}

\begin{abstract}

{\bf
The evolution of satellite galaxies is shaped by their constant interaction with the circum galactic medium surrounding central galaxies, which in turn may be affected by gas and energy ejected from the central supermassive black hole \cite{Cicone14,Woo16,Pillepich18,Nelson19,Oppenheimer20,Davies20}. However, the nature of this coupling between black holes and galaxies is highly debated \cite{Harrison18,Dashyan19,Veilleux20} and observational evidence remains scarce \cite{Cheung16,MN18b}. Here we report an analysis of archival data on 124,163 satellite galaxies in the potential wells of 29,631 dark matter halos with masses between 10$^{12}$ and $10^{14}$ solar masses. We find that quiescent satellites are relatively less frequent along the minor axis of their central galaxies. This observation might appear counterintuitive as black hole activity is expected to eject mass and energy preferentially in the direction of the minor axis of the host galaxy. However, we show that the observed signal results precisely from the ejective nature of black hole feedback in massive halos, as active galactic nuclei-powered outflows clear out the circumgalactic medium, reducing the ram pressure and thus preserving star formation. This interpretation is supported by the IllustrisTNG suite of cosmological numerical simulations, where a similar modulation is observed even though the sub-grid implementation of black hole feedback is effectively isotropic. Our results provide compelling observational evidence for the role of black holes in regulating galaxy evolution over spatial scales differing by several orders of magnitude.
}

\end{abstract}

\begin{document}

\flushbottom
\maketitle

\begin{figure} 
    \begin{center}
    \includegraphics[height=6.5cm]{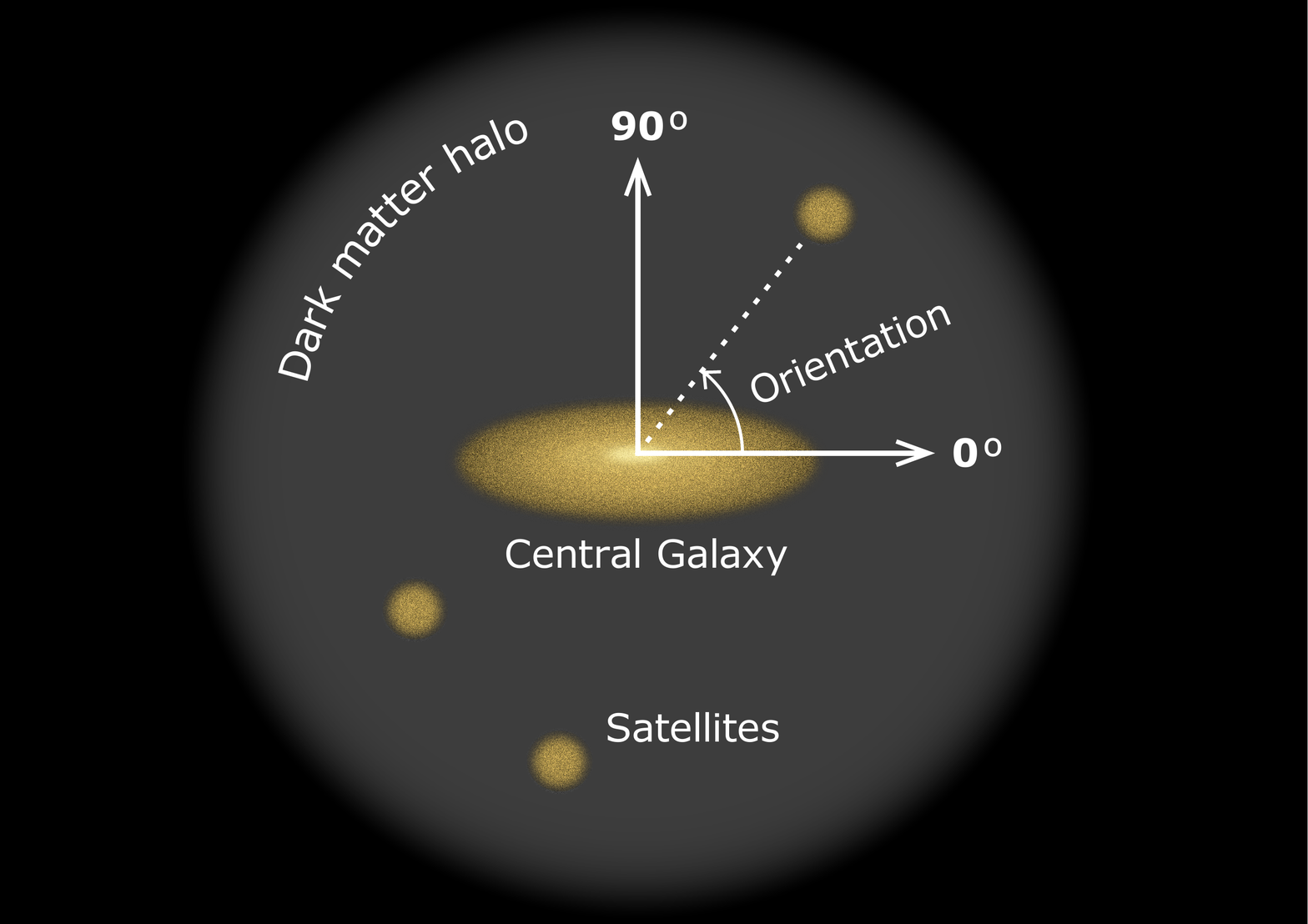}
    \end{center}
    \caption{{\bf Orientation of satellite galaxies around centrals.} For a given dark matter halo with a central galaxy, we define the orientation of each satellite as the difference between the position angle of the central's major axis and the position angle of the satellite with respect to the central, as indicated in the figure. A satellite located along the major axis of the central would have an orientation of 0\textdegree \ (or 180\textdegree/360\textdegree); conversely, a satellite located along the minor axis would have an orientation of 90\textdegree \ (or 270\textdegree).}
    \label{fig:1}
\end{figure}

We use catalogs of groups and clusters\cite{Tempel14} to identify satellite galaxies in the Sloan Digital Sky Survey (SDSS) \cite{SDSS10} data. For each satellite, its star formation rate (SFR) and stellar mass are measured \cite{Kauffmann03,Brinchmann04} (see Methods for details). We proceed to characterize how satellites are distributed around their centrals in the plane of the sky. As shown in Fig.~\ref{fig:1}, the orientation of a given satellite galaxy can be defined as the difference between the photometric position angle of the central's major axis and the position angle of the satellite with respect to the central. We retrieve the position angle of the central's major axis from the SDSS imaging pipeline \cite{Stoughton02} and we use sky coordinates to measure the position angle between centrals and satellites. Note that a galaxy does not need to be late type or of disky shape for a major axis to be defined, as in fact the great majority of the central massive galaxies in our SDSS sample are ellipticals. With this definition, a satellite located along the major axis of its central would have an orientation equal to 0\textdegree \ (or 180\textdegree/360\textdegree) and, conversely, a satellite located along the minor axis would have an orientation of 90\textdegree \ (or 270\textdegree). 

\begin{figure*} 
    \begin{center}
    \includegraphics[height=6.5cm]{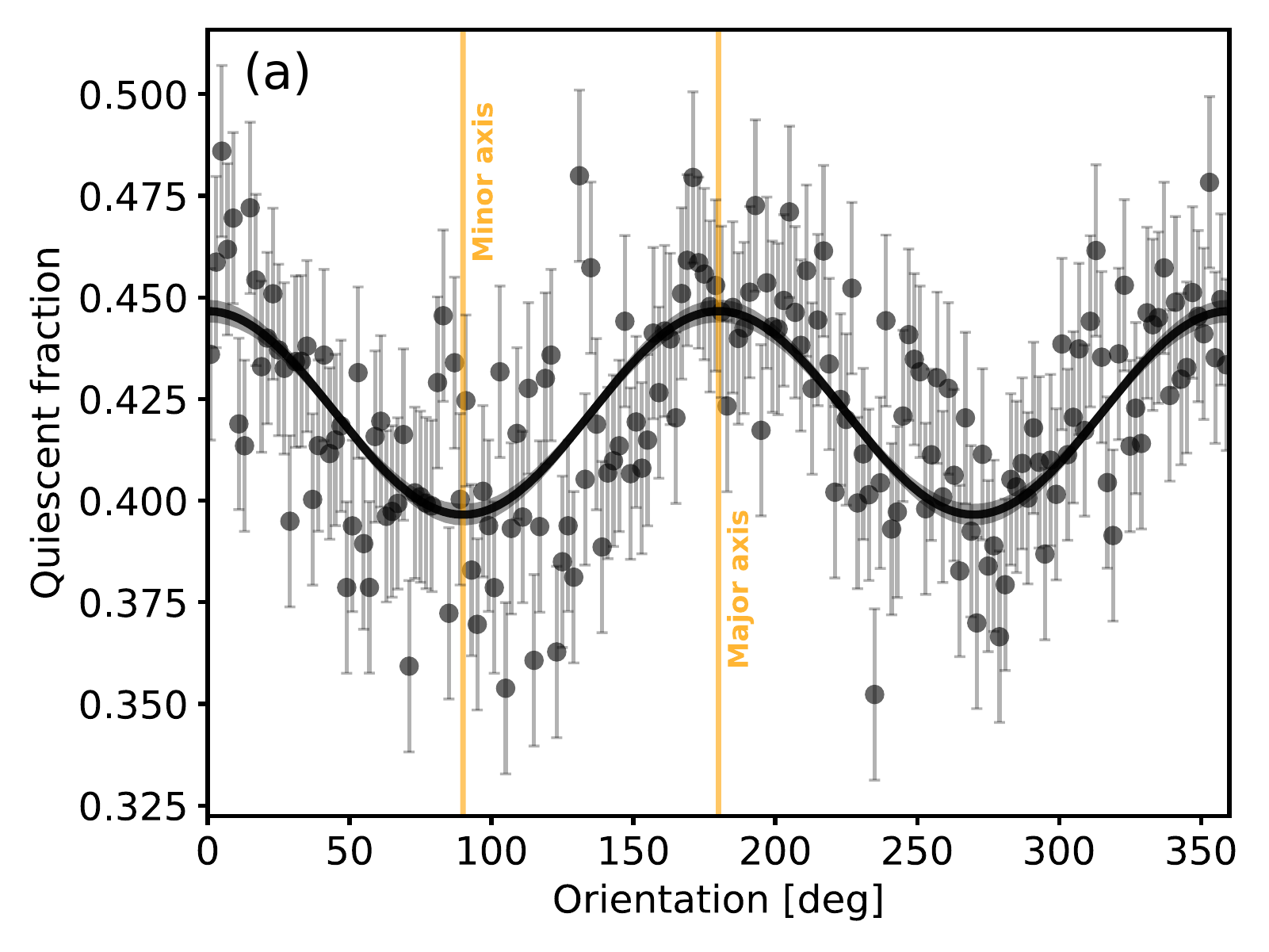}
    \includegraphics[height=6.5cm]{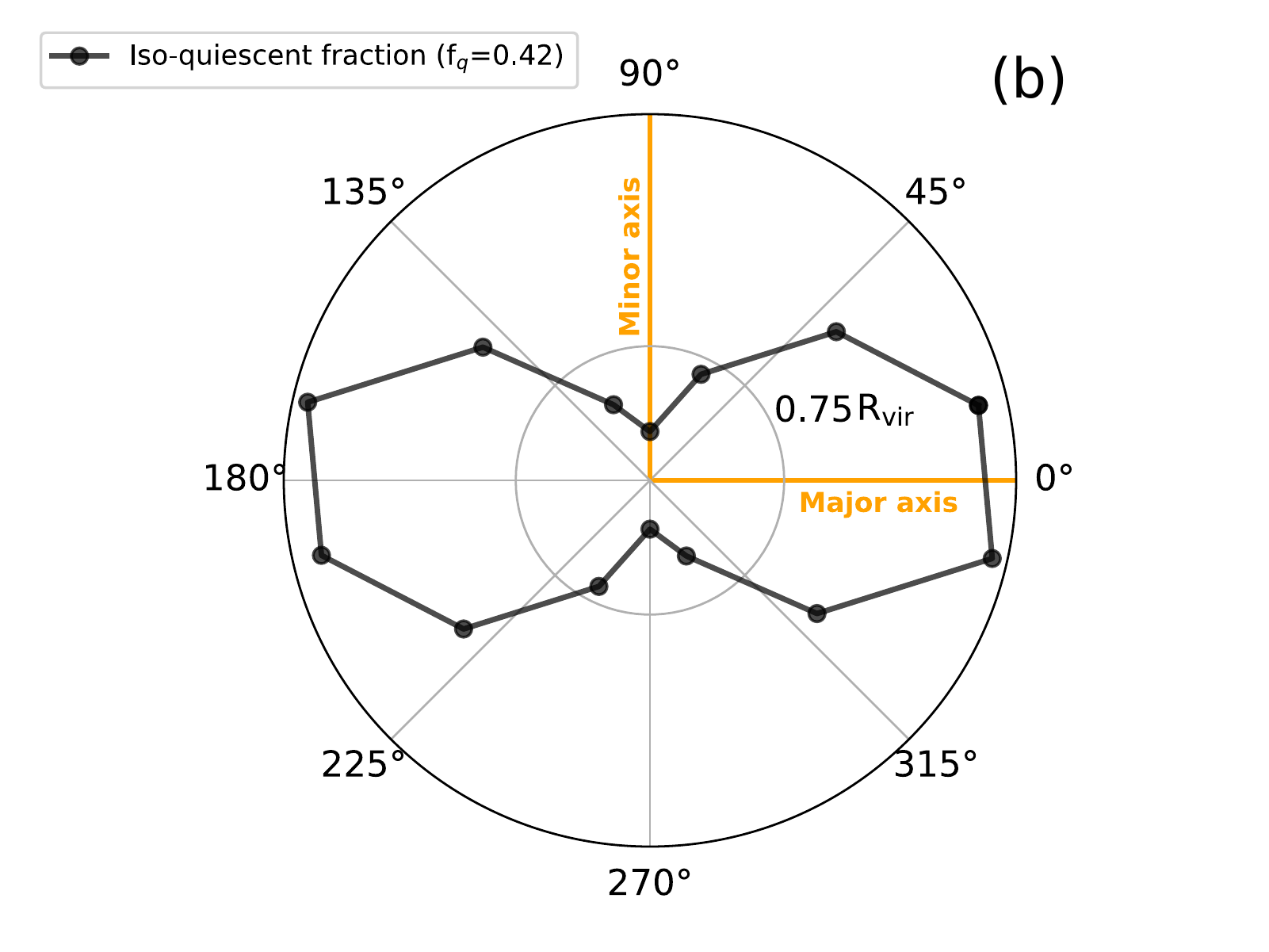}
    \end{center}
    \caption{{\bf Anisotropic distribution of quiescent satellite galaxies in SDSS.} Panel (a) shows how the fraction of quiescent galaxies depends on the orientation with respect to the central's major axis. Quiescent galaxies are relatively enhanced along the direction of the major axis and conversely, relatively deficient in the direction of the minor axis. The signal is well represented by a cosine function with an amplitude of 0.025 (black solid line), and the shaded area represents the 1$\sigma$ confidence interval ($\pm0.001$). Panel (b) represents the cluster-centric distance (normalized to the virial radius) at which the quiescent fraction equates 0.42 (the average across the adopted SDSS sample), as a function of orientation. The shape of the iso-quiescent contour demonstrates that satellite galaxies along the major axis of the central are quenched further out from the center of the cluster than those satellites located along the minor axis.}
    \label{fig:2}
\end{figure*}

Given this metric for the angular distribution of satellites galaxies and the ancillary SFR and stellar mass measurements, Fig.~\ref{fig:2} presents our main results. Panel (a) shows how the fraction of quiescent satellites depends on their orientation with respect to the central galaxy.  Each point indicates the fraction of quiescent galaxies among all satellites observed in a given orientation bin, regardless of the properties of the hosting halo. In practice this means that every group and cluster in our sample can contribute to each data point, depending on the location of its satellites. It is evident that  the fraction of quiescent satellites is maximal along the major axis of the central galaxy and minimal along the minor axis. In order to quantitatively assess the significance of this modulation, we fit the observed signal with a cosine function with three free parameters: the median quiescent fraction, the amplitude of the modulation, and a re-scaling of the assumed error to account for any source of uncertainty. The modulation is well represented by a cosine function with an amplitude of 0.025$\pm0.001$ on top of a 0.421$\pm0.001$ average quiescent fraction. Complementary, panel (b) shows the iso-quiescent fraction contour (f$_q$=0.42, the average of our sample) as a function of cluster-centric distance (normalized to the virial radius) and orientation angle. Along the major axis of the central galaxy, satellites are preferentially quenched at larger cluster-centric distances, while satellites located in the direction of the minor axis survive in relatively larger numbers as star-forming objects to much closer distances. Note that both panels in Fig.~\ref{fig:2} are obtained by stacking every group and cluster in our sample into a single pseudo-cluster, where the major axis of every central galaxy is aligned in the same direction. Note also that, in principle, the signal should be fully symmetrical around the 0\textdegree-90\textdegree range.

We have tested that this anisotropic modulation in the fraction of quiescent satellites, is statistically robust and exhibits a number of interesting features. As noted above the significance of a positive amplitude is well beyond 3$\sigma$ ($0.025\pm0.001$). Moreover, if we randomize the position angle of each central galaxy and we measure again the amplitude of the signal, we find in this case no modulation at all. At the same time, the signal is also robust against the assumed functional form for the surface brightness distribution while measuring the position angle of the central galaxy. The results shown in Fig.~\ref{fig:2} are thus robust and truly dependent on the orientation of satellites. We also find that the amplitude of the signal increases for more massive centrals and for less massive satellites. Moreover, the signal becomes stronger for satellites closer to the center of their host halo. Interestingly, the signal also depends on the mass of the black hole in the center of a halo: at fixed halo mass, the amplitude of the observed signal is stronger if the black hole hosted by the central galaxy is more massive. All these additional tests are presented in the Methods section.

\begin{figure} 
    \begin{center}
    \includegraphics[height=6.5cm]{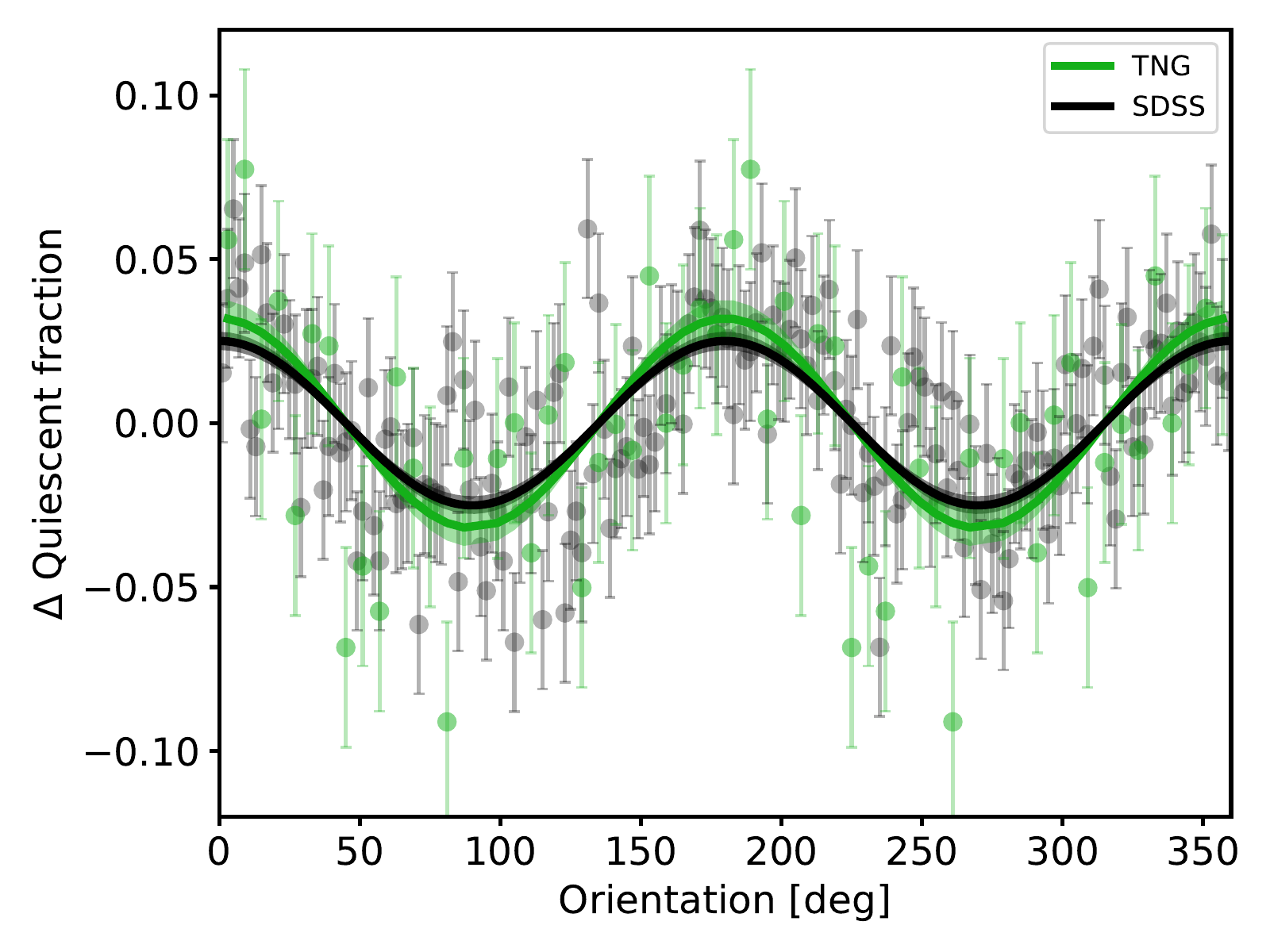}
    \end{center}
    \caption{{\bf SDSS vs IllustrisTNG.} Black symbols and lines are the SDSS observed data as in Fig.~\ref{fig:2}. Green symbols show the anisotropic distribution of quiescent satellites as measured in the IllustrisTNG 100Mpc volume cosmological numerical simulation. In the latter the amplitude (0.032) is similar to that observed in SDSS, and the  green shaded area indicates the 1$\sigma$ confidence interval ($\pm0.004$). Since the absolute fraction of quiescent galaxies  may differ between observed and simulated data, the mean quiescent fraction is subtracted to both datasets.  Due to the lower number of satellites, IllustrisTNG measurements correspond only to the 0\textdegree-180\textdegree interval. To show them along with the SDSS data we have assumed that the 180\textdegree-360\textdegree \ behaves exactly as from 0\textdegree-180\textdegree one.}
    \label{fig:3}
\end{figure}

To investigate the origin of the modulation, we make use of the IllustrisTNG suite of cosmological numerical simulations \cite{TNG}, where we find a similar behavior in the fraction of quiescent satellites. Fig.~\ref{fig:3} is based on the outcome of the TNG100 run, using high-realism synthetic SDSS-like images\cite{Vicente19} to measure the centrals' position angle. As in the SDSS data, the signal can be modeled by a cosine function with an amplitude of 0.032$\pm0.004$. Note that the selection function of SDSS satellites suffer from observational biases and completeness issues due to, e.g. fiber collisions. Hence, the absolute fraction of quiescent fractions might differ between SDSS and TNG100\cite{Martina20b}, particularly in groups and clusters. Therefore, in Fig.~\ref{fig:3} we have subtracted the average quiescent fraction of both datasets. Furthermore, we have verified that the signal is in place in IllustrisTNG irrespective of whether the photometric position angle or the intrinsic stellar angular momentum of the centrals are used to identify their major axes.

Two classes of distinct evolutionary mechanisms may be at the origin of the anisotropic distribution of quiescent satellites around central galaxies that we observe both in the SDSS data and in the IllustrisTNG cosmological numerical simulation. 

First, this could be the manifestation of a large-scale structure phenomenon, whereby satellites in groups and clusters form and evolve under different conditions than those in the field even before falling into the primary halos and thus, their properties and distribution can be affected by processes unrelated to their current group/cluster environment\cite{Fujita_2004}.

Second, this could be the result of a (host) halo phenomenon, i.e. emerging because of the very interaction between satellite galaxies and their host halo. This could be for example due to energetic feedback from e.g. the super-massive black holes that reside at the center of massive halos, which in turn can have an impact beyond the central galaxy itself, modulating also the evolution of the surrounding satellites \cite{Kauffmann_2013} through the alteration of the physical properties of the intra cluster/group medium. 

In order to tests these two scenarios, we compare the signal measured in IllustrisTNG and shown in Fig.~\ref{fig:3} with the signal measured from the previous generation, Illustris cosmological simulation\cite{Nelson15}. For the purposes of this work, the most noticeable difference between these two simulations is that IllustrisTNG includes an improved treatment of active galactic nuclei feedback, in particular in the low accretion rate regime\cite{Weinberger17,Pillepich18}. Thus, by comparing the two simulations, we are controlling for large-scale structure effects, isolating the role of black hole feedback in shaping the observed signal. We find that the modulation in the quiescent fraction significantly differs between the two simulations, being larger in IllustrisTNG (see Methods). For the first Illustris simulation, results are actually consistent with no modulation at a $\sim2\sigma$ level (0.013$\pm 0.07$). Furthermore, within the IllustrisTNG galaxy population, the anisotropic signal is driven by satellites that quenched within their current host halo (i.e. were not quenched prior to infall due to, e.g., pre-processing) and by the star-forming population of satellite galaxies (see Methods).

We therefore conclude that the anisotropic behavior in the distribution of quiescent galaxies is a host halo phenomenon and, in particular, an observable manifestation of AGN feedback acting far beyond the extension of the central host galaxy\cite{Dave19}. This AGN-driven origin is particularly supported by the reported connection between the amplitude of the signal and the mass of the central black hole in the observed data, and more explicitly by the relation between energy injection and signal modulation in the simulations.  Additionally, in both SDSS and IllustrisTNG data, the anisotropic quenching signal is stronger around quiescent rather than star-forming centrals, further suggesting a connection to the activity of the super-massive black holes at the centers.

Moreover, it is consistent with our analysis of the IllustrisTNG sample where, in the explored halo mass range, most satellite galaxies become quiescent after infalling in their current halo and the anisotropic signal in the simulation is dominated by such galaxies. Furthermore, AGN feedback is expected to impact more prominently the halo gas in the vicinity of the center, and in the SDSS data we find that satellites at smaller galactocentric distances exhibit a larger anisotropic modulation of quiescence.

\begin{figure*}[!htpb]
    \begin{center}
    \includegraphics[height=6.5cm]{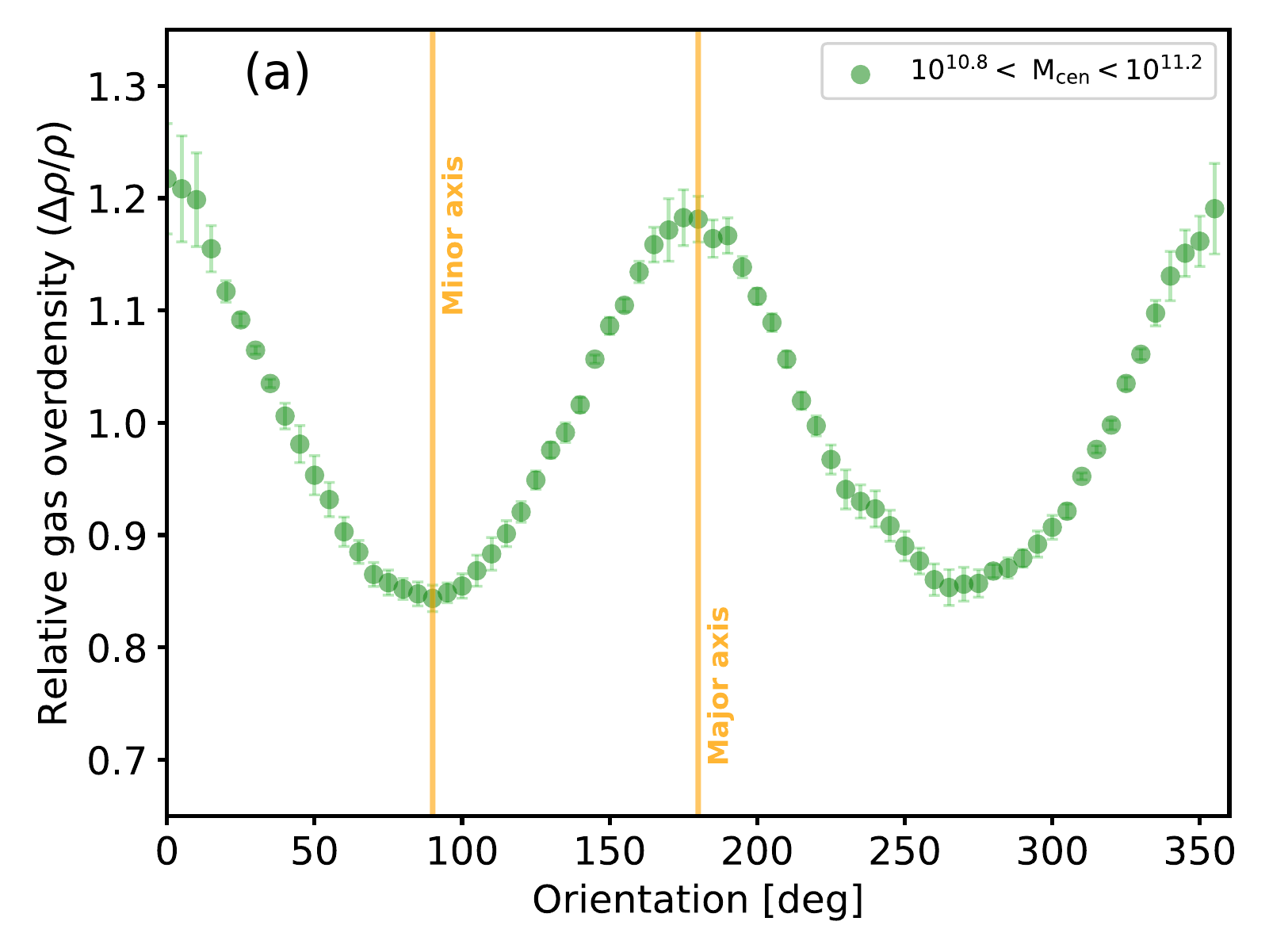}
    \includegraphics[height=6.5cm]{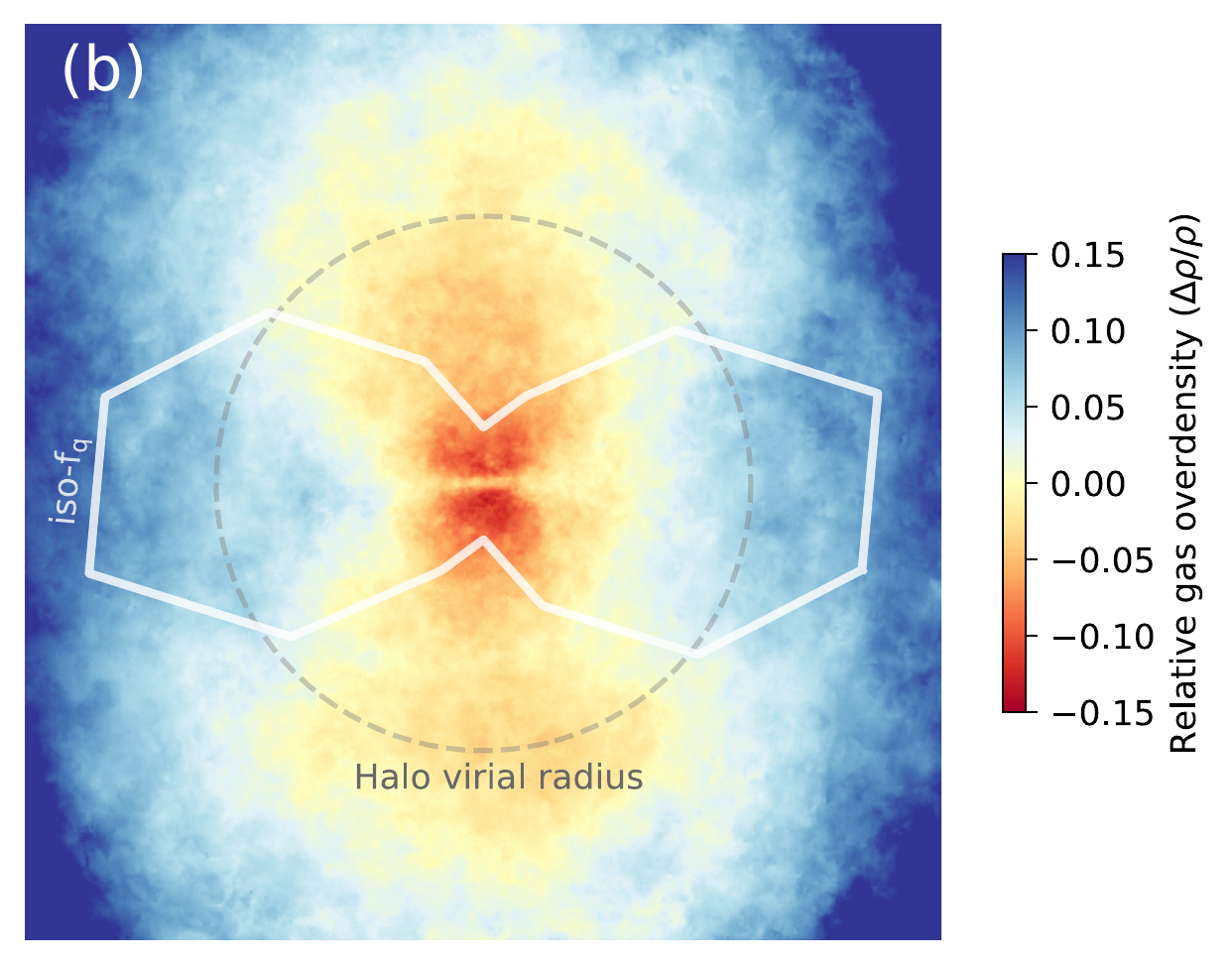}
    \end{center}
    \caption{{\bf Anisotropic CGM density in IllustrisTNG.} Panel (a) illustrates how the mean relative gas overdensity changes as a function of orientation for central galaxies in  the TNG100 simulation with stellar masses around the median of our SDSS sample ($M_\mathrm{cen}\sim 10^{11}$ \msun). Averaged within the virial radius, the relative gas overdensity exhibits a similar behaviour as that measured for the fraction of quiescent galaxies in Fig.~\ref{fig:2}, as the circumgalactic medium gas density is relatively lower along the minor axis of centrals.  Panel (b) shows a stacked average image of the gas overdensity (relative to its azimuthal average) over the same mass range, with the iso-quiescent fraction contour of Fig,~\ref{fig:1} overlaid in white. For reference, the size of the virial radius is also indicated (dashed gray circle).}
    \label{fig:4}
\end{figure*}

The relative lower fraction of quiescent satellites along the minor axis of central galaxies and its relation with AGN feedback may at first seem counterintuitive because the energy and mass radiated by AGN activity are expected to escape the central galaxy preferentially along that direction. However, we argue that it is precisely the ejection of energy along the minor axis what drives the observed signal, as AGN feedback carves low-density bubbles in the CGM surrounding the central galaxy\cite{McNamara07,Fabian12}. This is exemplified in Fig.~\ref{fig:4}, where we show how the average mass density of the CGM around central galaxies in IllustrisTNG displays exactly such anisotropic behavior, despite the fact that feedback processes are isotropic at the scale of energy injection. 
The interaction of  the outflows with the galactic gas and the inner CGM is also likely responsible for the observed geometry, as the outflowing gas would tend to follow the path of least resistance, decoupled from the direction of energy injection at the smallest scales. This is supported by the fact that in the IllustrisTNG simulations, energy injection from super-massive black holes does not have a preferred direction, and yet outflows appear bi-polar on large scales. In any case, the anisotropic distribution of the CGM around central galaxies shown in Fig.~\ref{fig:4}, a direct IllustrisTNG prediction for the proposed scenario, should be observable in X-rays with sufficient statistics.

All our findings support the idea that it is the interaction between satellites and the CGM, in turn modulated by AGN activity, that drives the observed signal. We suggest that, as satellite galaxies pass through these low-density regions, processes directly responsible for their quenching such as ram pressure stripping become less efficient, increasing the relative abundance of star-forming galaxies along the direction of the minor axis. This interpretation is consistent with our analysis of the SDSS data, as ram pressure stripping is expected to affect more severely low mass satellites hosted by more massive centrals as indeed observed \cite{Gunn72, Yun19}. We cannot in fact exclude a different scenario, whereby the star formation activity of the satellites is enhanced rather than their quenching suppressed: it is possible that star-formation is enhanced along the outflowing material\cite{Maiolino17}, further increasing the fraction of star-forming satellites in the direction of the minor axis.

\bibliography{bubbles_varXiv}

\section*{Acknowledgements}

IMN acknowledges support from grant PID2019-107427GB-C32 from The Spanish Ministry of Science and Innovation and from the Marie Sk\l odowska-Curie Individual {\it SPanD} Fellowship 702607. AP and MD acknowledge support by the Deutsche Forschungsgemeinschaft (DFG, German Research Foundation) -- Project-ID 138713538 -- SFB 881 (``The Milky Way System''), subproject A01.

\section*{Author contributions statement}
IMN and AP developed the original idea and characterized the signal in the observed and simulated data. DN measured the gas mass density distribution in IllustrisTNG and contributed to the early developement of the project. VRG generated the synthetic SDSS-like images based on IllustrisTNG data, and MD provided the information about the infalling time of satellites in IllustrisTNG. LH and VS contributed to the analysis and interpretation of the observed and simulated data. IMN and AP wrote the text, and all the co-authors contributed to refine and polish the final manuscript.

\section*{Author information}
Reprints and permissions information is available at www.nature.com/reprints

\section*{Data Availability Statement}
All data used in this work are publicly available through the Sloan Digital Sky Survey and the Illustris and IllustrisTNG public data releases.

\noindent
The authors do not have any competing financial interests

\noindent
Correspondence and requests for materials should be addressed to {\it imartin@iac.es}

\vspace{1cm}

\clearpage

\section*{Methods}

\setcounter{figure}{0}  

\section{Sample properties}

\subsection{Sloan Digital Sky Server data}
Galaxy groups and clusters are selected\cite{Tempel14} from the 10th Data Release of the Sloan Digital Sky survey\cite{SDSS10}. For each group and cluster, the Navarro-Frenk-White\cite{NFW} mass of the dark matter halo is provided, along with an estimation of the virial radius. In addition, the catalog also contains a list of satellites and identifies the central of each halo. We cross-match this catalog of groups and clusters with a catalog of SFR\cite{Brinchmann04} and stellar mass\cite{Kauffmann03} measurements based as well on spectro-photometric data from the SDSS. This matching of catalogs provides a final sample of 124,165 satellites with stellar masses in the M$_\star \sim 10^{8.2}$-$10^{11.5}$\msun \ range associated to 29,631 galaxy groups and clusters, hosted in halos with masses ranging from $M_\mathrm{halo}\sim 10^{11.7}$  \ to $M_\mathrm{halo}\sim 10^{14.5}$ \msun  \ and with central galaxy mass in the M$_\star \sim 10^{9.3}$-$10^{11.9}$\msun \ range, at a median redshift of $z = 0.08$. The typical distance of satellites with respect to central galaxies ranges from $\sim$50 kpc to $\sim 900$ kpc. No cuts were imposed on the basis of the b/a axis ratios of central galaxies.

The separation between star-forming and quiescent satellites is based on their location with respect to the star formation main sequence (SFMS). In particular, we find that the SFMS in our sample of satellites is well-fitted\cite{MN19} by $\log \rm{SFR} = 0.75 \log M_\star - 7.5$. As described in the main text, given this definition of the SFMS and having stellar masses and SFR measurements for each satellite, we label as star-forming any satellite departing less than 1 dex from the SFMS. Conversely, a satellite is labeled as quiescent if its SFR is below the SFMS by more than 1 dex. 

In order to define the orientation of a satellite (see Fig.~\ref{fig:1}) the position angle (PA) of the central's major axis has to be estimated. Thus, for each central in our sample, we use photometric information retrieved from the SDSS imaging pipeline \cite{Stoughton02}. Since two measurements of the PA are given, depending on whether the central is better fit by an exponential or a de Vaucouleurs\cite{dV} profile, we choose the parametrization with the highest likelihood given by the {\it lnLExp} (exponential) or {\it lnLDeV} (de Vaucouleurs) keywords. To increase the robustness of the assumed quantities, we average likelihoods and 
PA over $g$, $r$, and $i$ bands. 

\subsection{IllustrisTNG simulation data}

We focus our analysis of the IllutrisTNG cosmological magneto-hydrodynamical simulations on the 100 Mpc volume run, known as TNG100, for which synthetic SDSS-like images and photometric measurements have been made in a forward-modeling fashion \cite{Vicente19}. For simulated galaxies with stellar mass above M$_\star = 10^{9.5}$ \msun, we generate an SDSS-like synthetic image using the SKIRT radiative transfer code \cite{Baes11,Camps13} and perform a two-dimensional Sérsic fit using the {\it statmorph} code \cite{Vicente19}. We then use the best-fitting PA in the same way as in the SDSS data. This approach has two main advantages. First, image fitting is done on the projected X-Y plane of the simulation, which naturally emulates the distribution of orientations expected for central galaxies in the SDSS data. Second, a S\'ersic fit generalizes the exponential and de Vaucouleurs profiles used in the SDSS imaging pipeline, ensuring a relatively fair comparison between observations and simulations. Images were synthesized using the snapshot 99 corresponding to redshift $z=0$ and we adopt the same metric and definition of quiescent and star-forming as in the SDSS data:  note that the SFMS of TNG100 is well consistent with observational constraints at low redshifts\cite{martina19}.

Flagging out those centrals with unsuccessful S\'ersic fits, we make use of a total of 880 halos (and centrals) in the host halo mass range of $M_\mathrm{halo}\sim 10^{12}$ -$10^{14.2}$  and of 8,552 satellites with stellar masses ranging from M$_\star \sim 10^{8}$ to  M$_\star \sim 10^{11.6}$ \msun. At the resolution of TNG100, the less massive satellites would contain of the order of $\sim100$ stellar particles \cite{Martina20b}. For central galaxies, stellar masses range from M$_\star \sim 10^{9.5}$ to $10^{12.2}$ \msun. Details on galaxy and halo identification can be found in the IllustrisTNG presentation papers\cite{Pillepich18,TNG}. Despite the large cosmological volume of 100 Mpc, the number of objects is significantly lower than in SDSS, although enough to reveal the presence of a modulation in the number of quiescent satellites (see Fig.~\ref{fig:3}). 

It is worth emphasizing that the goal of comparing SDSS and IllustrisTNG data is to understand the origin of the observed modulation in the fraction of quiescent galaxies, and not to provide an even-handed comparison between observed and simulated properties of galaxies in absolute terms:  previous works have shown that the IllustrisTNG galaxy population is in good agreement with SDSS results, e.g. in terms of galaxy colors, global and small-scale stellar morphologies, SFRs, and quenched fractions\cite{Nelson18, Huertas-Company19,Zanisi20,Martina20b}. In particular, the IllustrisTNG outcome is in striking quantitative agreement in comparison to SDSS data (differences smaller than $<5-10$ percentage points) for global central and satellite quenched fractions at $z = 0.1$ and, in terms of satellites quenched fractions as a function of halocentric distance, for intermediate, group-mass scale hosts ($10^{13}-10^{14}$ \msun)\cite{Martina20b}, which dominate the host mass distribution of both the SDSS and IllustrisTNG samples adopted here. We rely on those findings to elect IllustrisTNG as our simulation counterpart. At the same time, we note that an exact matching between the SDSS and IllustrisTNG samples is not critical for the purpose of this work, as we can get insights by focusing on the relative effects of the satellite locations and by marginalizing over the possibly-different absolute fractions of quiescent galaxies. On  the other hand, matching exactly the galaxy samples would lead to an even lower number of available satellites, hampering the reliability of the comparison.

\section{Signal characterization}

In this section we detail the most meaningful tests done, both in  the observed and simulated galaxy datasets, to understand the origin of the observed modulation in the number of quiescent satellites.

\subsection{Fitting the observed signal} 

\begin{extfig*} 
    \begin{center}
    \includegraphics[height=9cm]{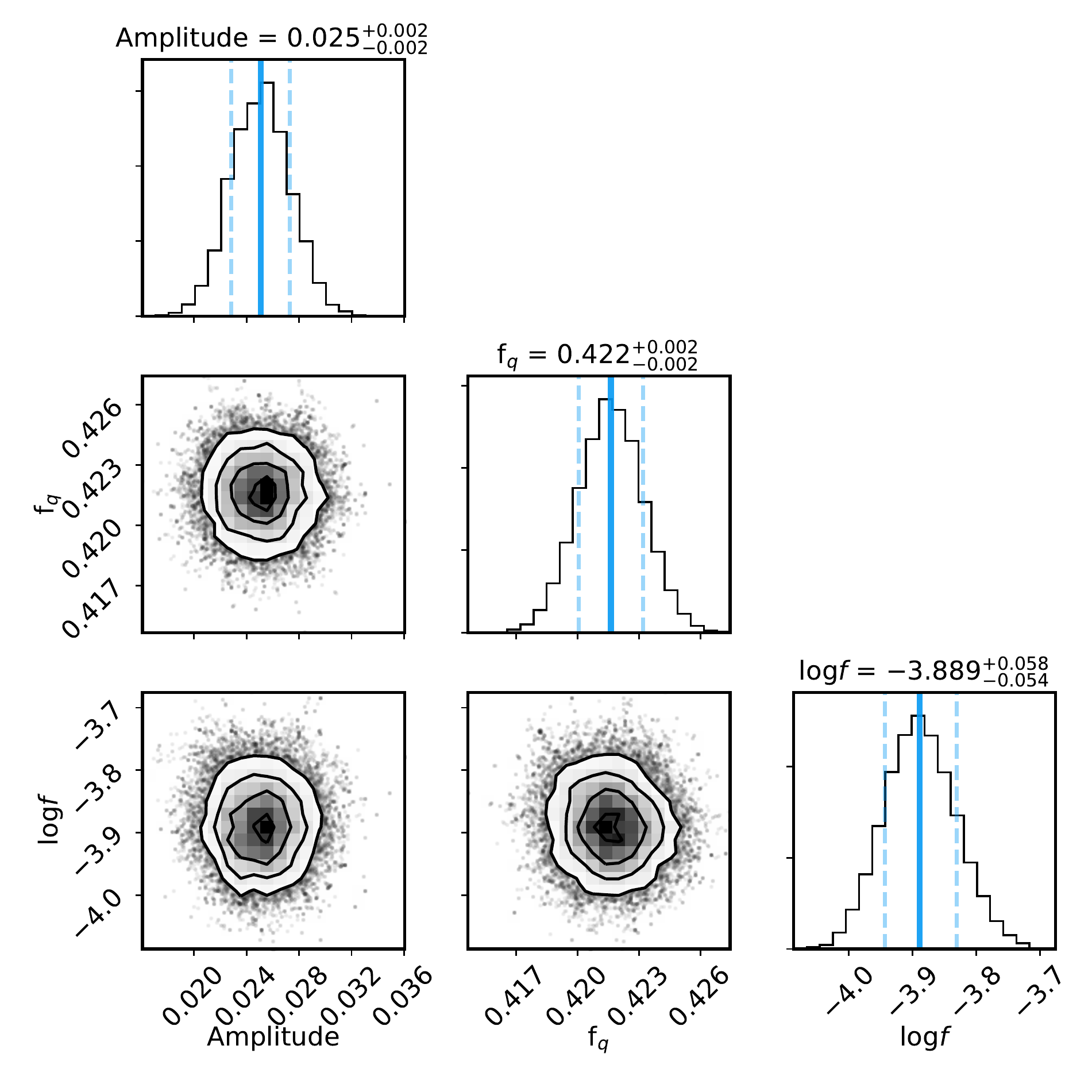}
    \end{center}
    \caption{{\bf SDSS posterior distributions for the best-fitting description of the angular modulation of satellite quiescence.} We fit the observed data with a cosine function with three free parameters, the average quiescent fraction ($f_q$), the amplitude of the modulation, and a re-scaling term for the expected error $f$. Posteriors are well-behaved and allowed us to reject the null-hypothesis at a $\sim 6 \sigma$ level. Blue solid vertical lines indicate the best-fitting values and the dashed ones the 1$\sigma$ confidence interval.
    }
    \label{fig_methods:post}
\end{extfig*}

With no other motivation than quantifying the angular dependence of the fraction of quiescent galaxies, we fit the observed signal with a cosine function. In practice, we use a Bayesian MCMC sampler\cite{emcee} to evaluate the following likelihood function

\begin{equation*}
    \ln p(\mathrm{f_q} | \theta, a, b, f) = - \frac{1}{2} \sum_i \left[ \frac{(\mathrm{f_q}_i - a - b \cos 2\theta_i)^2}{s^2_i} + \ln (2 \pi s_i^2)\right], 
\end{equation*}

\noindent
where $\mathrm{f_q}_i$ is the observed fraction of quiescent galaxies at the $\theta_i$ orientation, $a$ is the median quiescent fraction, and $b$ is the amplitude of the modulation. The error term is given by $s_i^2 = \sigma_i^2 + f^2$, where $\sigma$ is the estimated error, in our case estimated by bootstrapping, and $f$ is the re-scaling term. 

This Bayesian framework allows us to explore the full posterior distribution and to naturally tests the null hypothesis (i.e. {\it Is the signal consistent with an amplitude equal to zero?}). In Extended Data Fig.~\ref{fig_methods:post} we show the posterior distribution for the best-fitting solution of the SDSS data (Fig.~\ref{fig:2}). With these assumptions, we can reject the null hypothesis at a $\sim 6 \sigma$ level. 

Note that error bars shown in all figures represent the combined $s^2 = \sigma^2 + f^2$ uncertainty, and, for practical reasons, we fit for logarithmic re-scaling of the error $\ln f$. The inclusion of this additional error term $f$ allows us to account for the various sources of error that might affect the observed data, such as the uncertainty on the estimated PA of the central galaxy, on the stellar mass and star formation rate measurements of the satellites, and on their stochastic distribution around centrals.

\subsection{Robustness of the signal against the assumed position angle}

Since the orientation of satellites strongly depend on the assumed PA of the central galaxy, we test the robustness of our results against errors and systematics on the PA determination. First, we test the sensitivity of the observed signal to the functional form assumed for the surface brightness distribution of central galaxies. This is motivated by the fact that the light profile of galaxies is not accurately fitted by either a single exponential or de Vaucouleurs function as assumed by the SDSS photometric pipeline. Similarly to Fig.~\ref{fig:2}, panel (a) in Extended Data Fig.~\ref{fig_methods:PA} shows the modulation in the fraction of quiescent satellites, but this time intentionally assuming the worst-fitting functional form according to SDSS. For example, if the SDSS photometric pipeline indicates that a galaxy is better fitted by an exponential profile, we selected the PA derived using a de Vaucouleurs, and vice versa. As it becomes obvious from panel (a) in Extended Data Fig.~\ref{fig_methods:PA}, the observed modulation in the fraction of quiescent satellites is insensitive to the functional formed used to fit the brightness profile of the central galaxy. 

\begin{extfig} 
    \begin{center}
    \includegraphics[height=6.5cm]{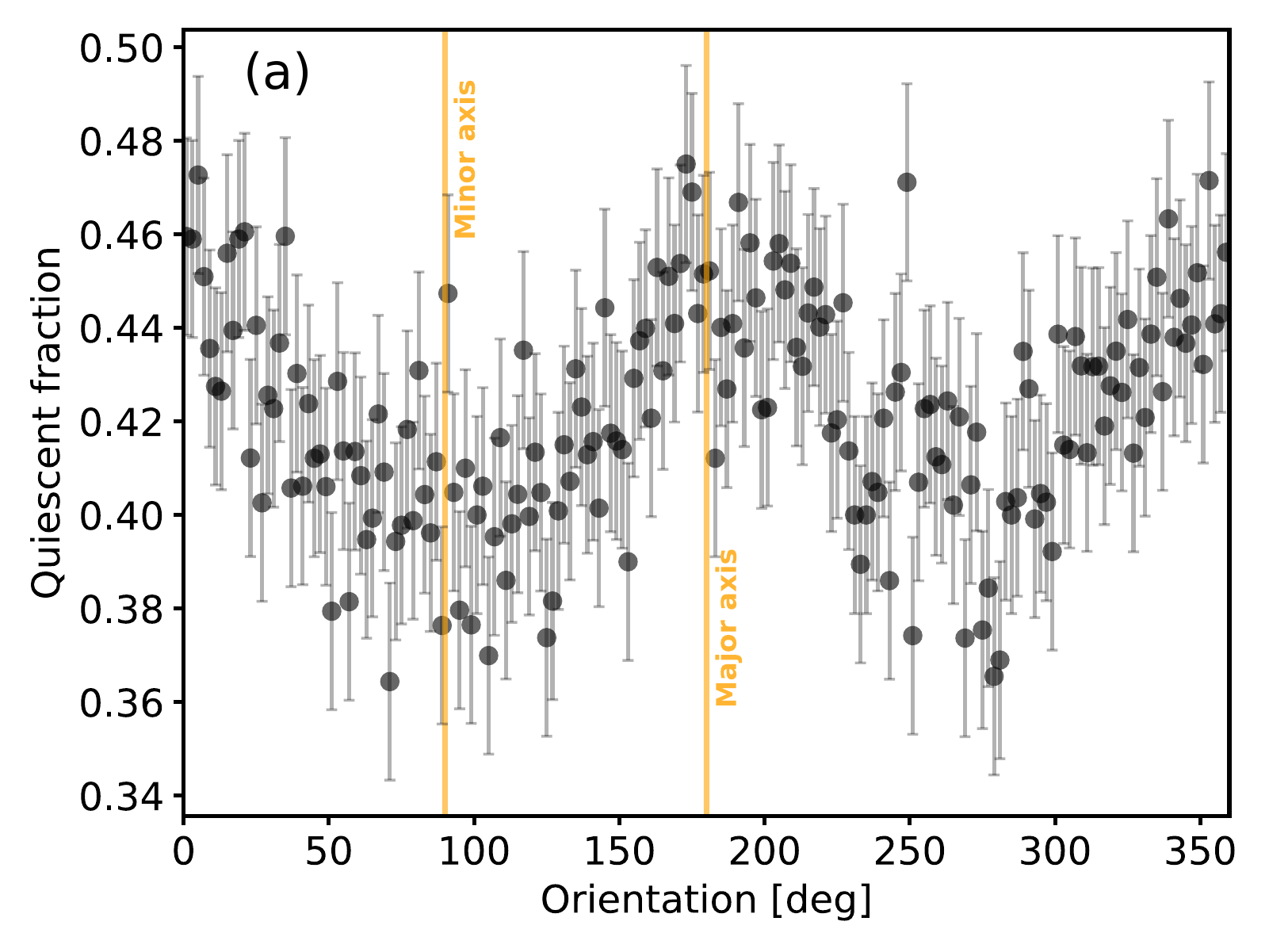}
    \includegraphics[height=6.5cm]{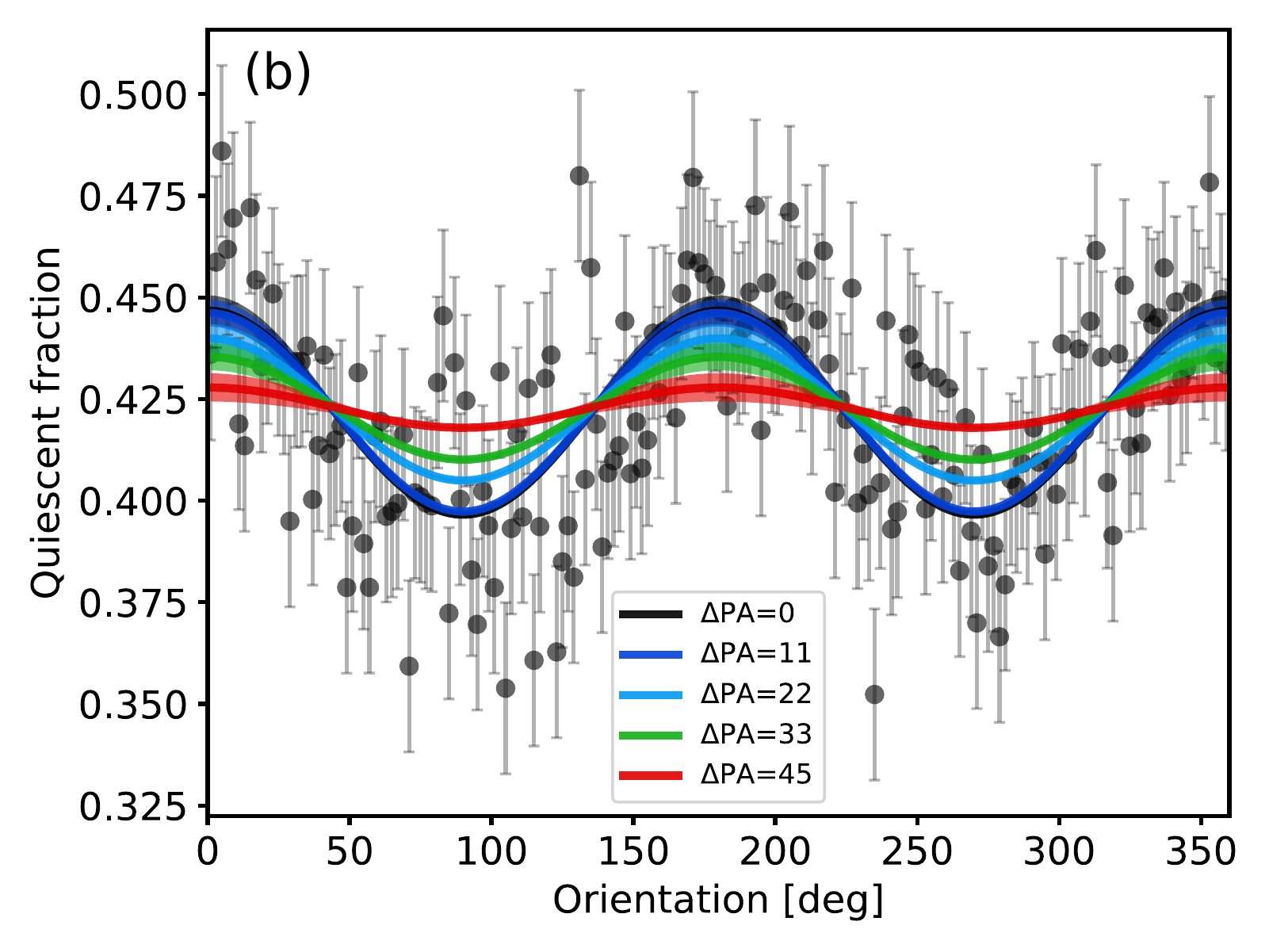}

    \includegraphics[height=3.4cm]{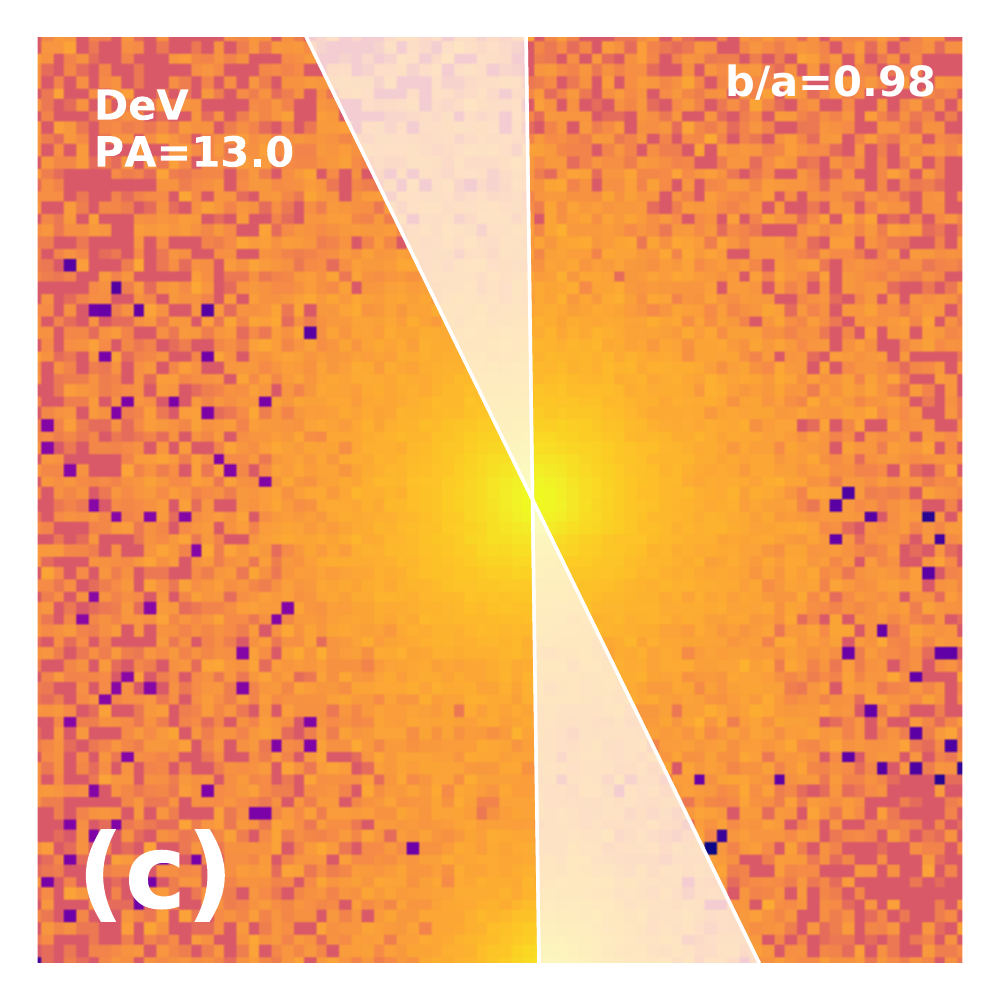}
    \includegraphics[height=3.4cm]{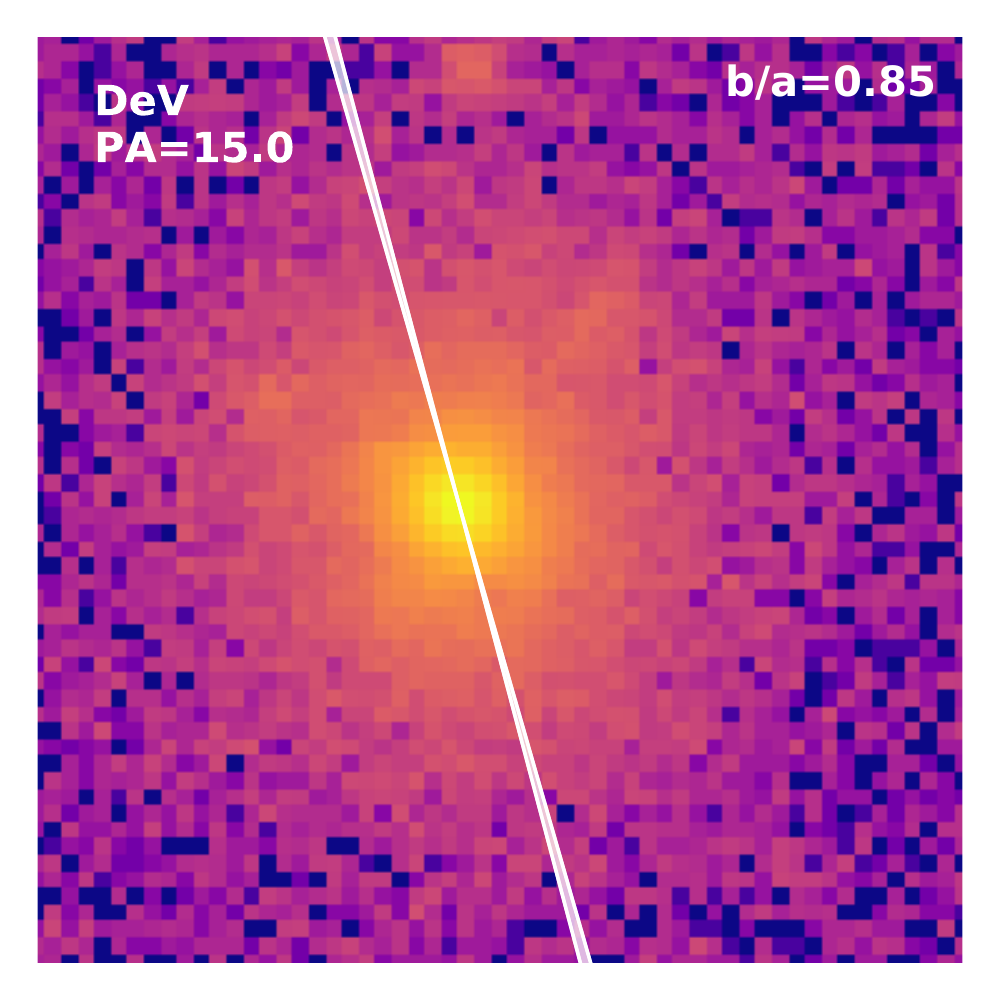}
    \includegraphics[height=3.4cm]{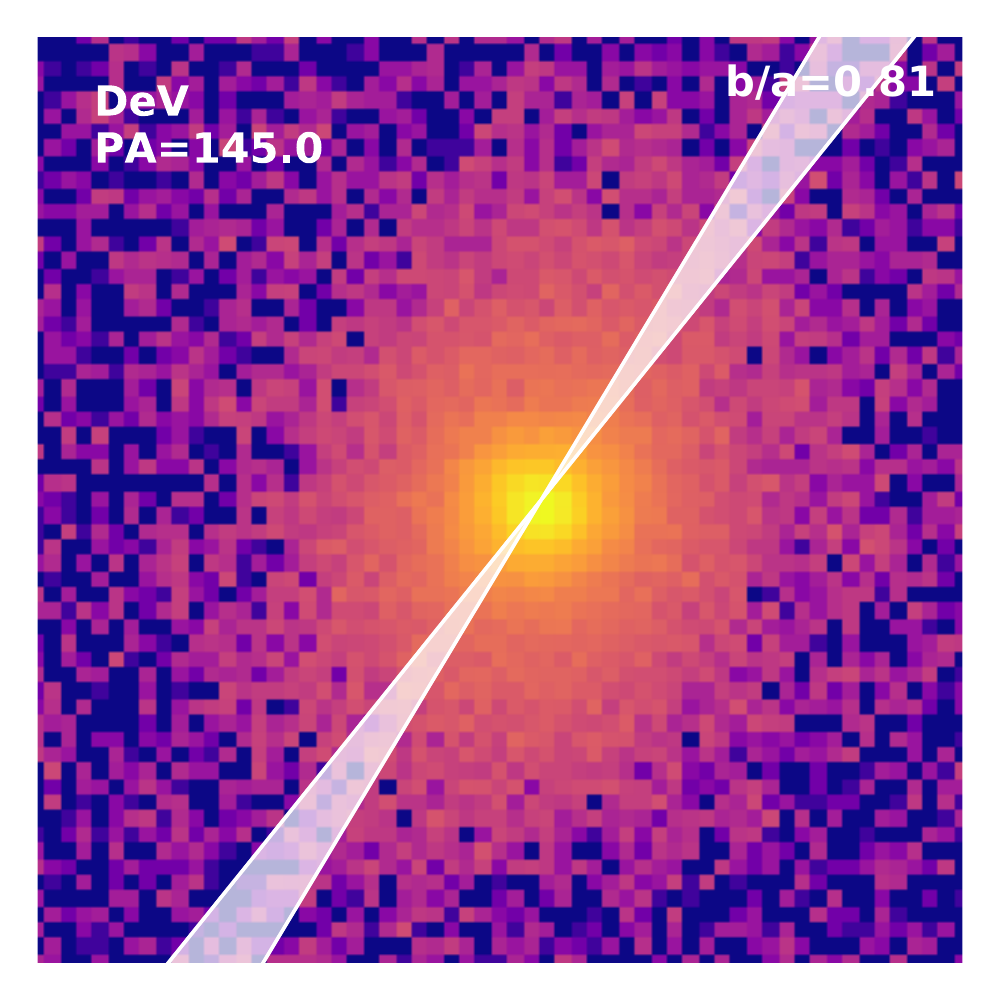}
    \includegraphics[height=3.4cm]{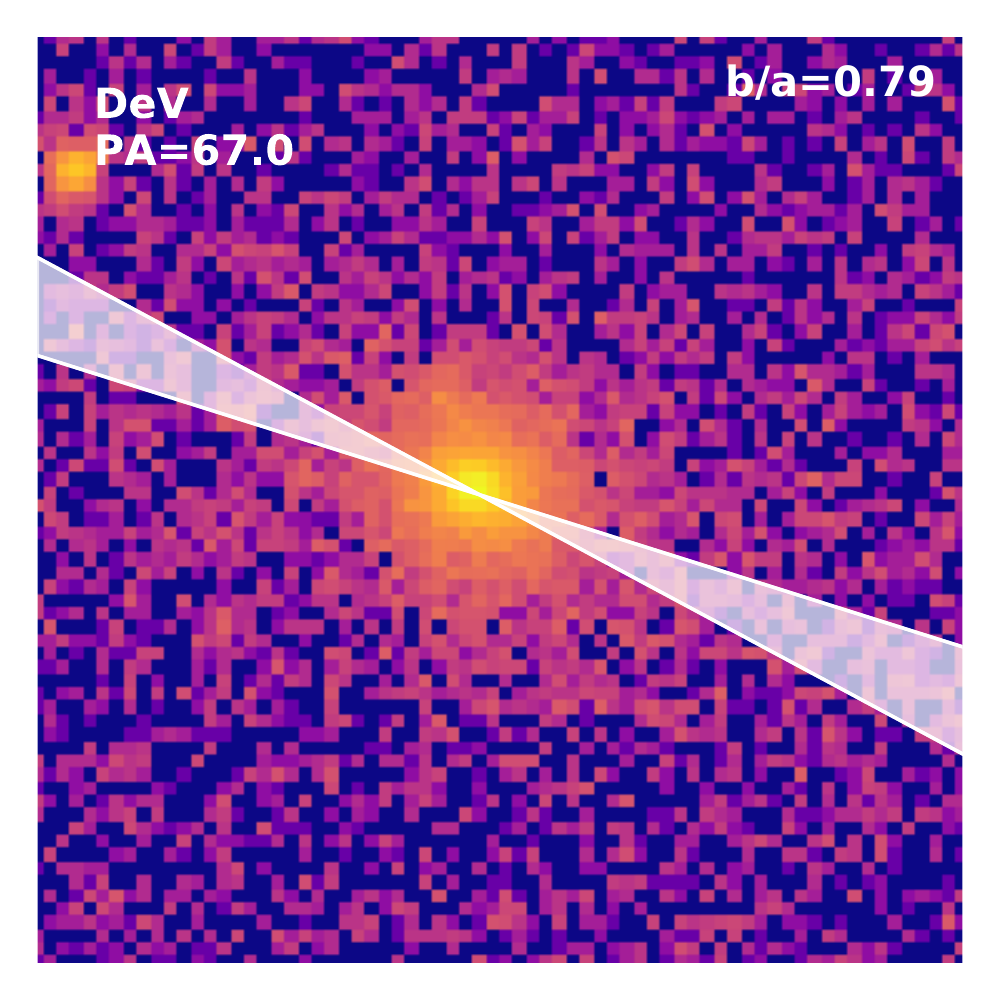}
    \includegraphics[height=3.4cm]{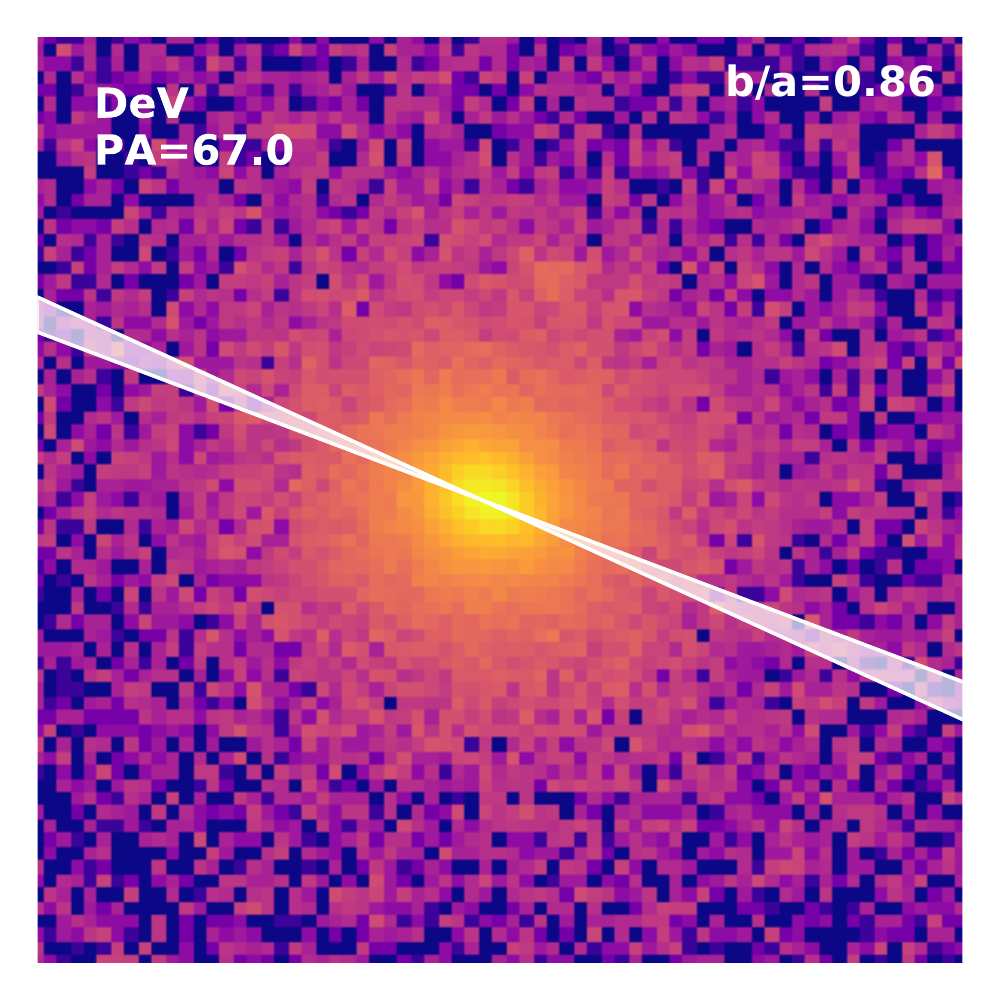}

    \includegraphics[height=3.4cm]{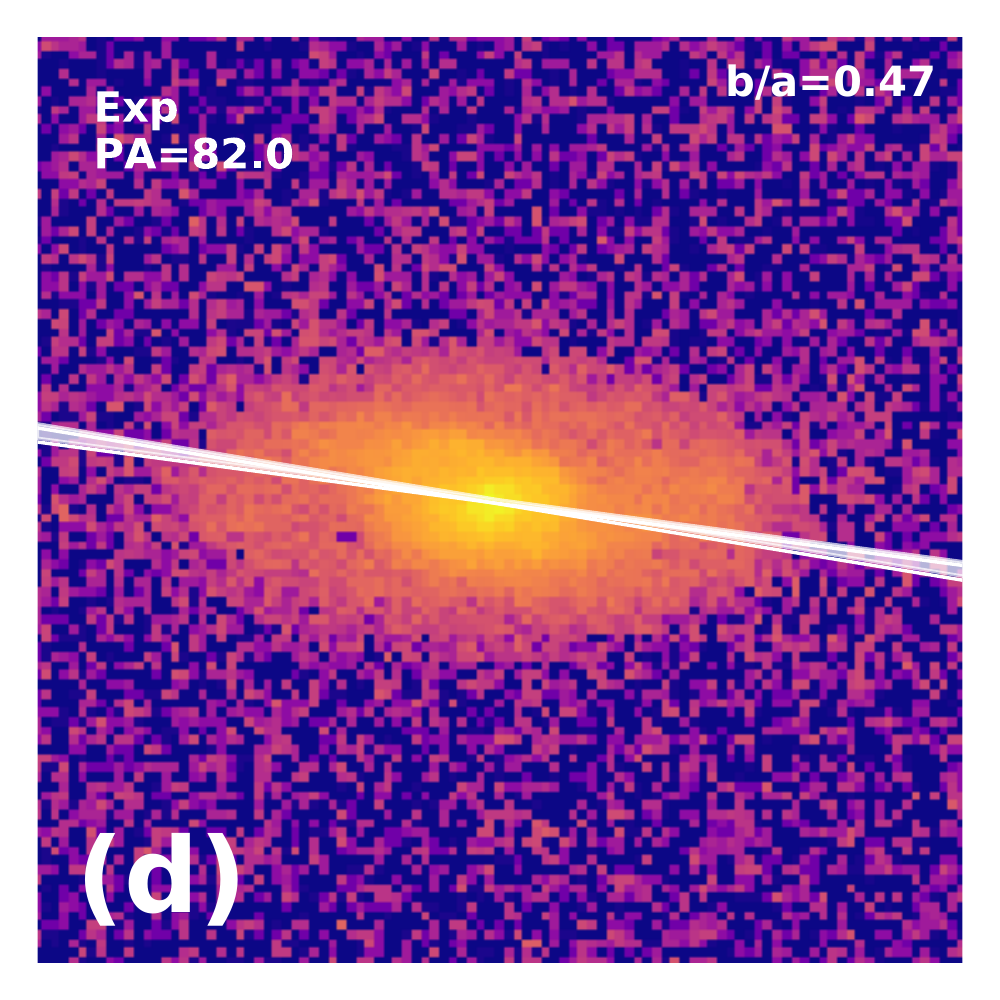}
    \includegraphics[height=3.4cm]{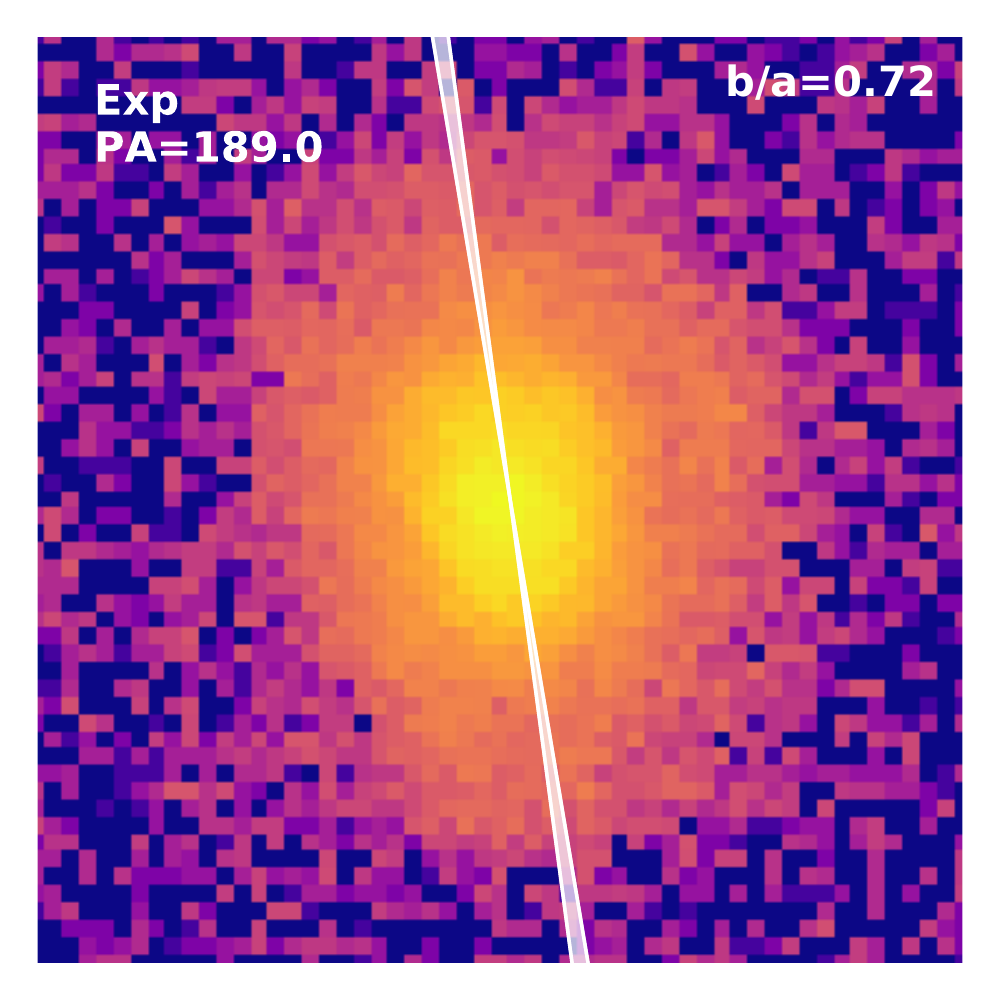}
    \includegraphics[height=3.4cm]{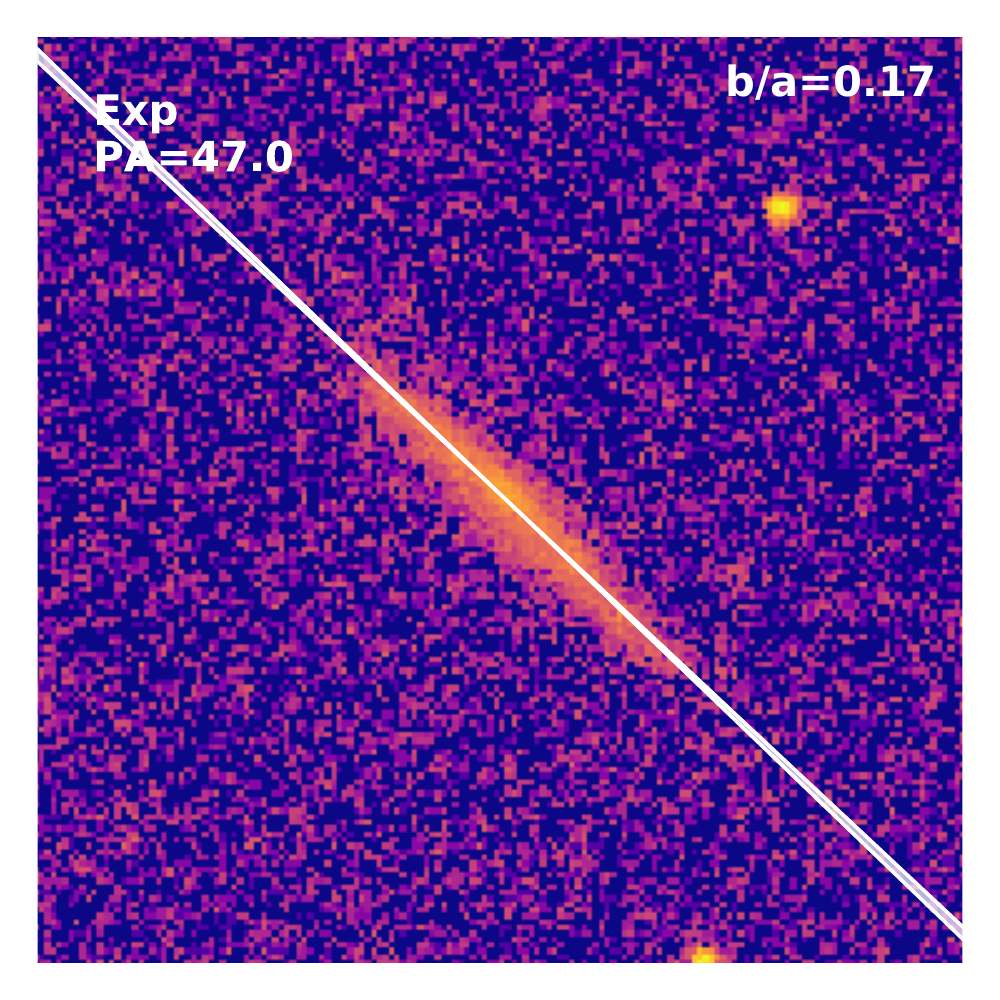}
    \includegraphics[height=3.4cm]{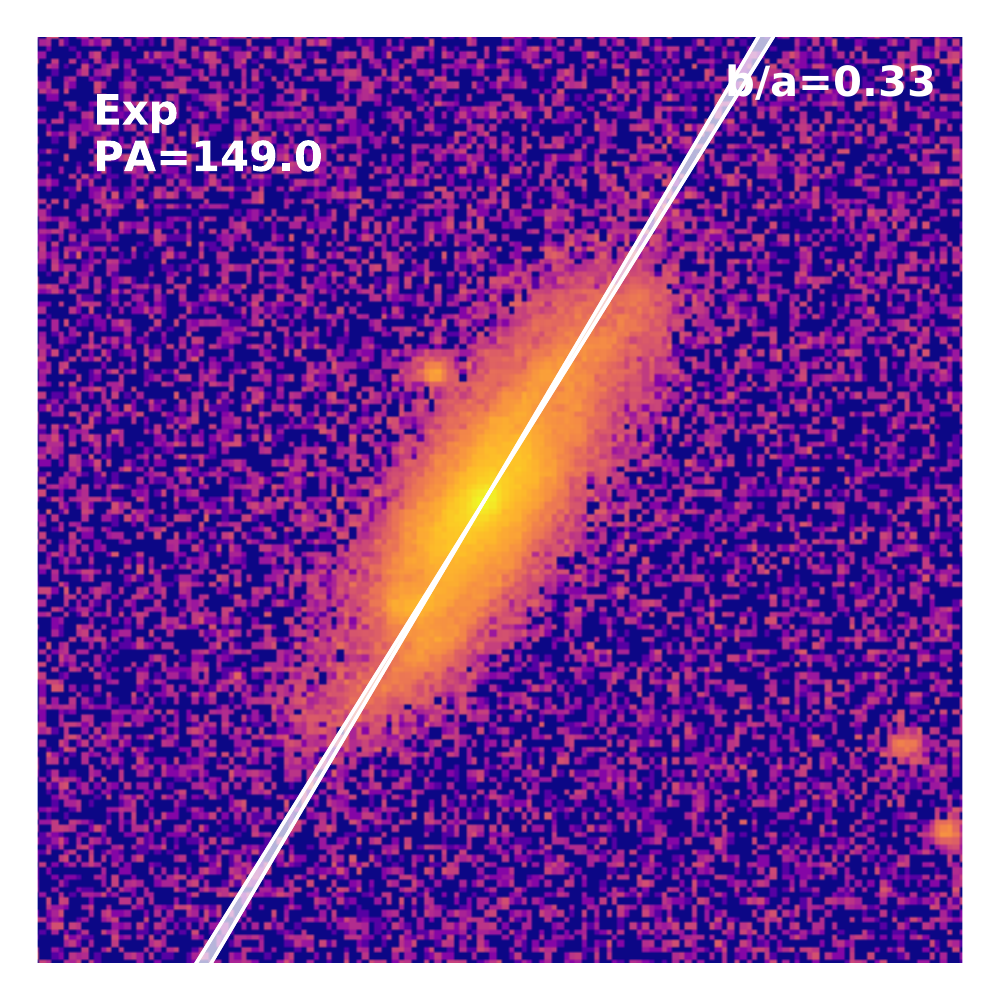}
    \includegraphics[height=3.4cm]{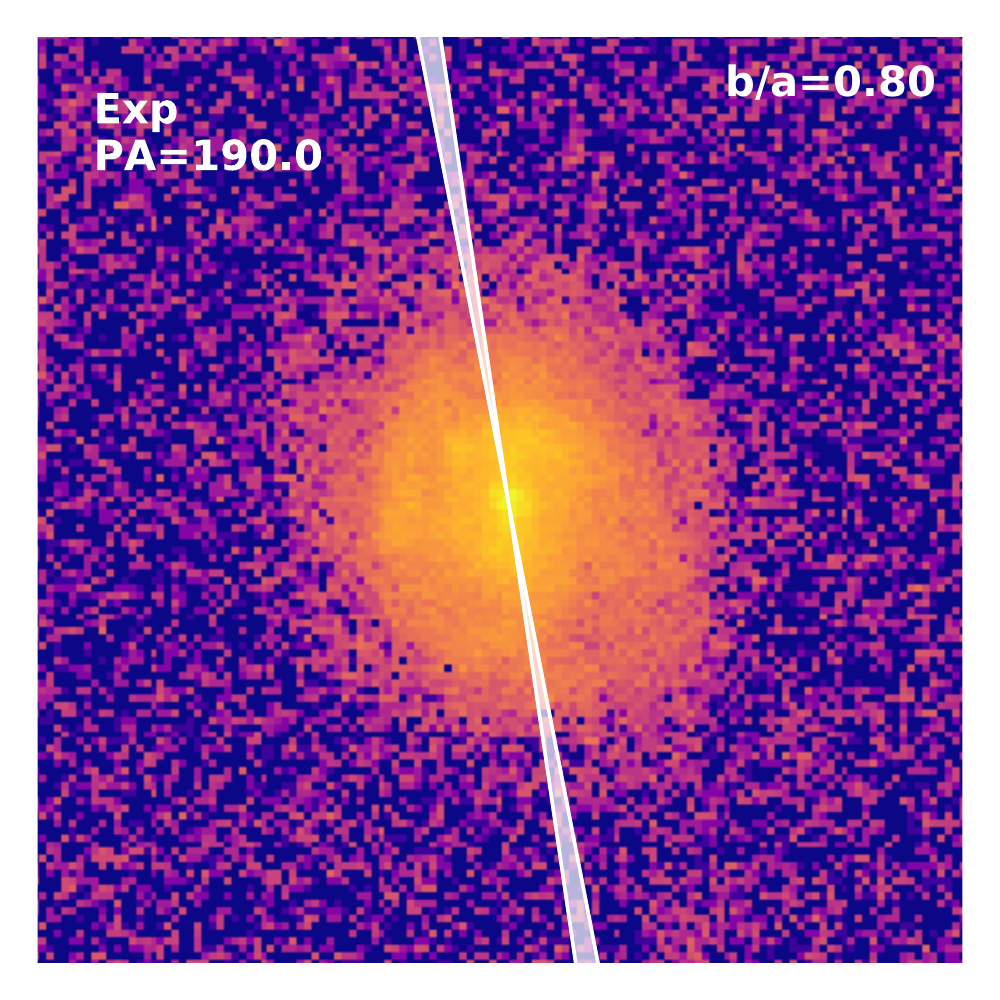}
    \end{center}
    \caption{{\bf Sensitivity of the SDSS signal to PA uncertainties.} Panel (a) shows the fraction of SDSS quiescent galaxies as a function of the orientation based on the worst-fitting functional form (de Vaucouleurs vs. exponential) according to the SDSS photometric pipeline. The stability of the signal demonstrates that our results are robust against the photometric fitting procedure. In panel (b), colored curves indicate the best-fitting solution for SDSS data obtained while randomly perturbating the PA of the central galaxy by $\Delta$PA. For reference, black symbols and curves are the same as in Fig.~\ref{fig:2}. A clear modulation in the fraction of quiescent galaxies is observed even for $\Delta$PA$\sim$30\textdegree, which is an order of magnitude larger than the expected error on the individual PAs. Panels (c) and (d) show show the SDSS $g$-band images of galaxies best-fitted by a de Vaucouleurs (top row) and an exponential profile (bottom row), with the PA uncertainty indicated by the white shaded area. The adopted PA is indicated in the top left corner of each image.
    }
    \label{fig_methods:PA}
\end{extfig}

Additionally, we also test if errors in the determination of the central's PAs could have a significant impact on the observed signal. This is done by perturbing the PA of each central galaxy by a factor $\delta$PA, drawn from a normal distribution $\mathcal{N}(0,\Delta \mathrm{PA}^2)$. By doing this we practically investigate the effect that a typical error of $\Delta \mathrm{PA}$ would have on the recovered signal, and it is shown in panel (b) of Extended Data Fig.~\ref{fig_methods:PA}. For errors on the individual PAs of up to 30\textdegree, there is still a clear modulation in the fraction of quiescent satellites. Note that the expected typical error in our sample of galaxies\cite{Simard11} is of the order of $\sim2$\textdegree, an order of magnitude smaller than the extreme test shown in Extended Data Fig.~\ref{fig_methods:PA}. Hence, we conclude that systematics and uncertainties related to the PA measurements  based on the available photometric data in our sample of SDSS galaxies do not significantly affect our findings and thus our conclusions. For illustrative purposes, panels (c) and (d) in Extended Data Fig.~\ref{fig_methods:PA} show examples of galaxies with de Vaucouleurs and an exponential light profiles, respectively. The uncertainty in the assumed position angle is indicated with the white dashed area.

As a final test, we repeated our analysis but completely randomizing the PA of each central galaxy. In this case, the orientation of satellites becomes meaningless and therefore it is expected that the signal disappears. This is indeed the case as shown in panel (a) of Extended Data Fig.~\ref{fig_methods:random}. Panel (b) in the same figure shows the posterior distribution, demonstrating that the amplitude of the signal vanishes when the centrals' PAs are randomized.

\begin{extfig} 
    \begin{center}
    \includegraphics[height=6.5cm]{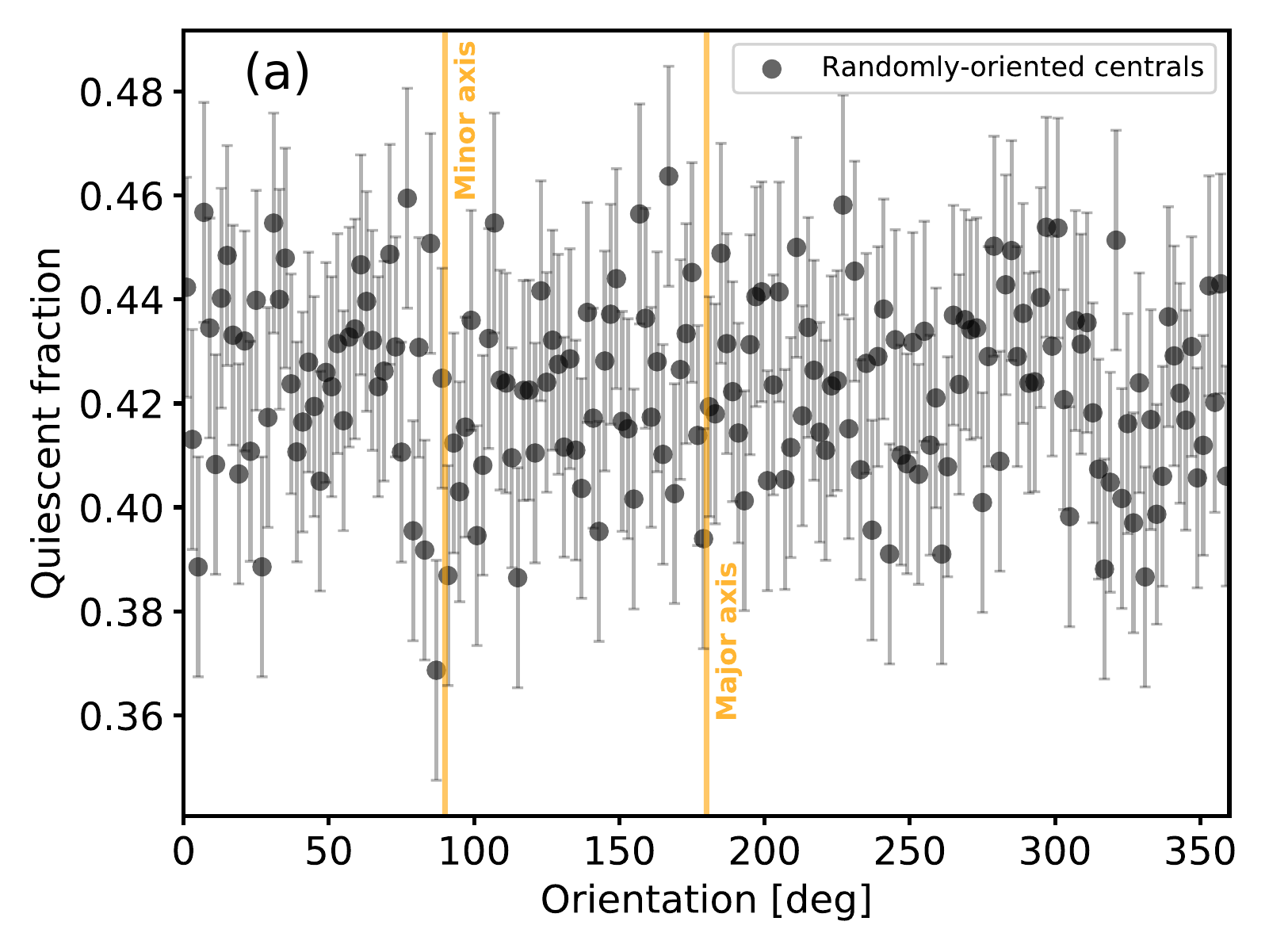}
    \includegraphics[height=6.5cm]{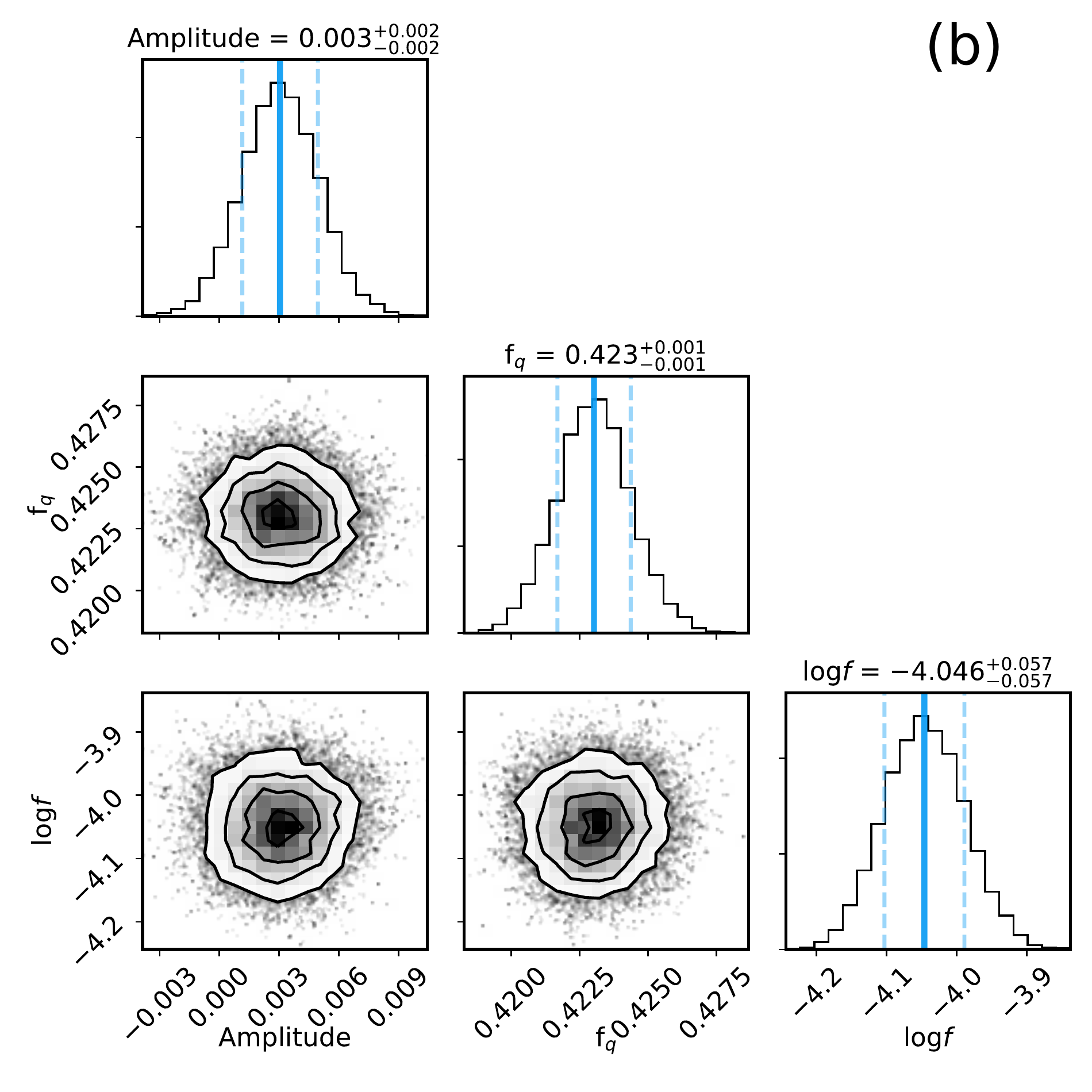}
    \end{center}
    \caption{{\bf Test with randomized position angles.} Panel (a) shows the fraction quiescent satellites in SDSS data after randomizing the PA of the central galaxies. As expected, no signal is recovered in this case. Panel (b) shows the posterior distributions for this test, where the modelled amplitude is consistent with no angular variation.
    }
    \label{fig_methods:random}
\end{extfig}

\subsection{Dependence on galaxy properties in SDSS data} 

The large number of objects in the SDSS sample of satellites allows us to investigate the dependence of the observed signal on different properties of centrals and satellite galaxies. Extended Data Fig.~\ref{fig_methods:1} summarizes the more meaningful trends we find in the SDSS data set. In particular,  in panel (a) we split our sample into satellites close to the center of their halos/centrals (R$_\mathrm{sat} < 0.5$R$_\mathrm{vir}$) and satellites farther out from the center (R$_\mathrm{sat} > 0.5$R$_\mathrm{vir}$). For those satellites closer to the center we find an amplitude of 0.038 $\pm 0.002$ in the modulation of the signal, while for those in the outskirts the modulation was weaker, with an amplitude of 0.023 $\pm 0.002$. 

\begin{extfig*} 
    \begin{center}
    \includegraphics[height=4.4cm]{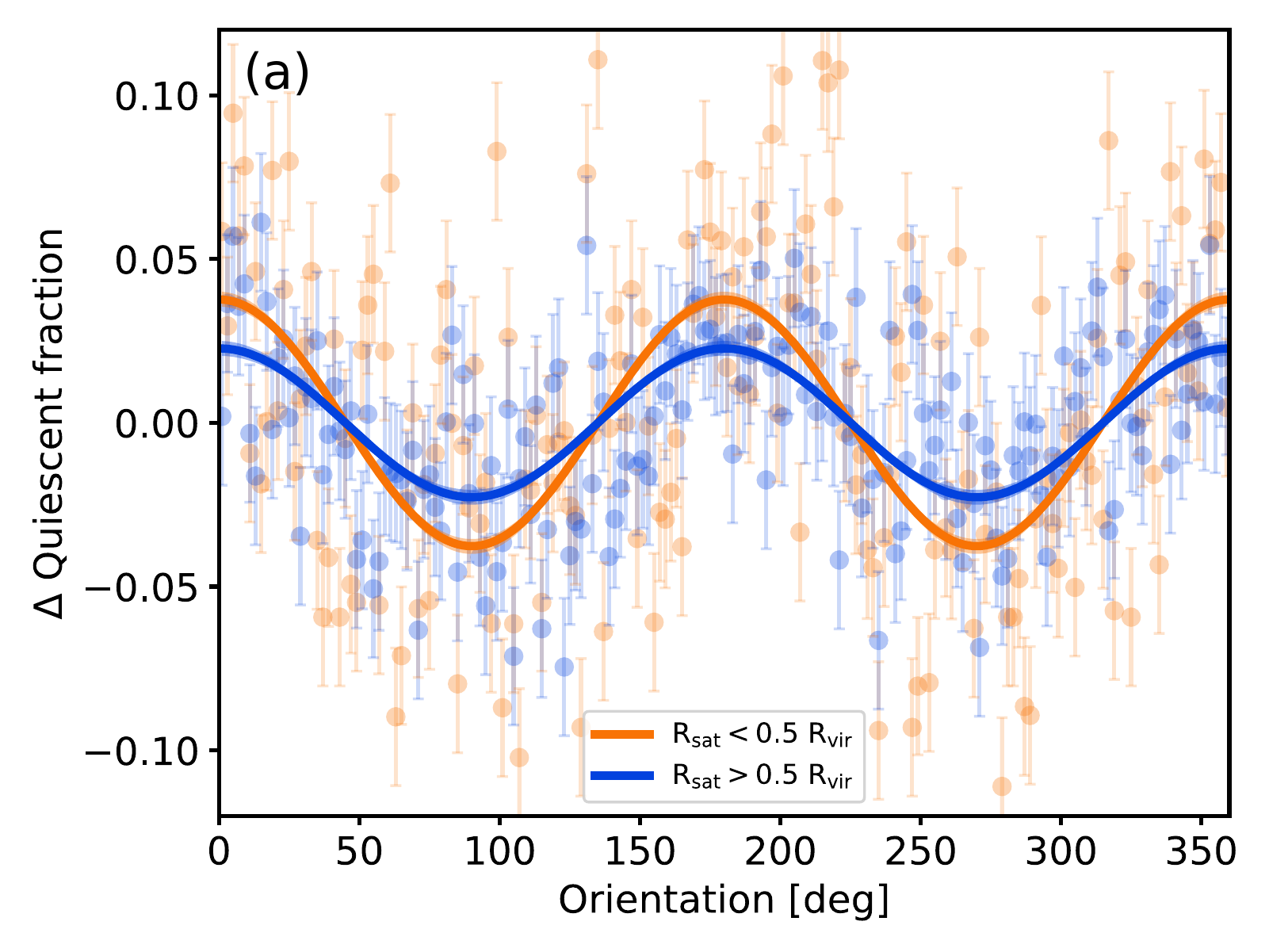}
    \includegraphics[height=4.4cm]{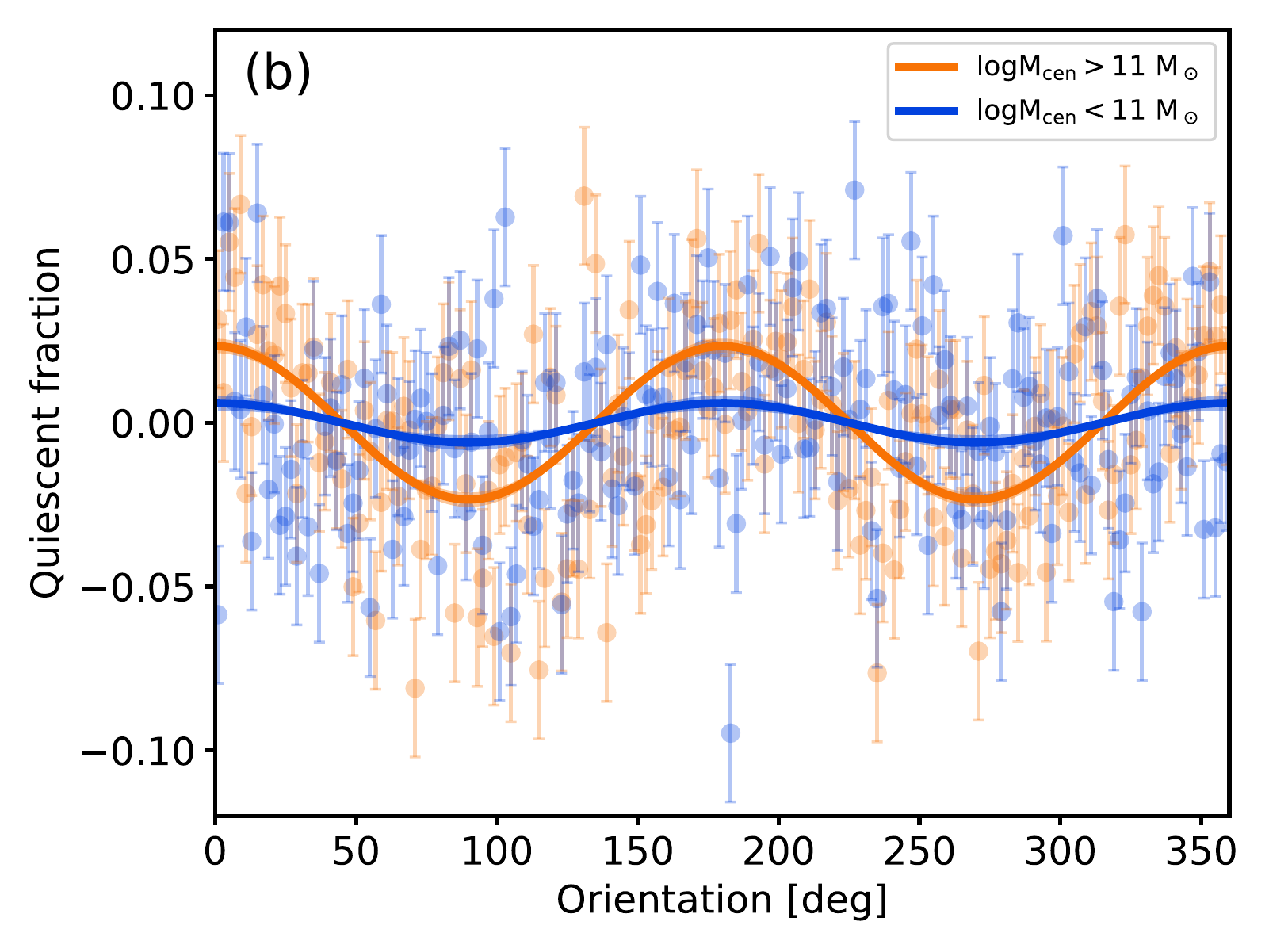}
    \includegraphics[height=4.4cm]{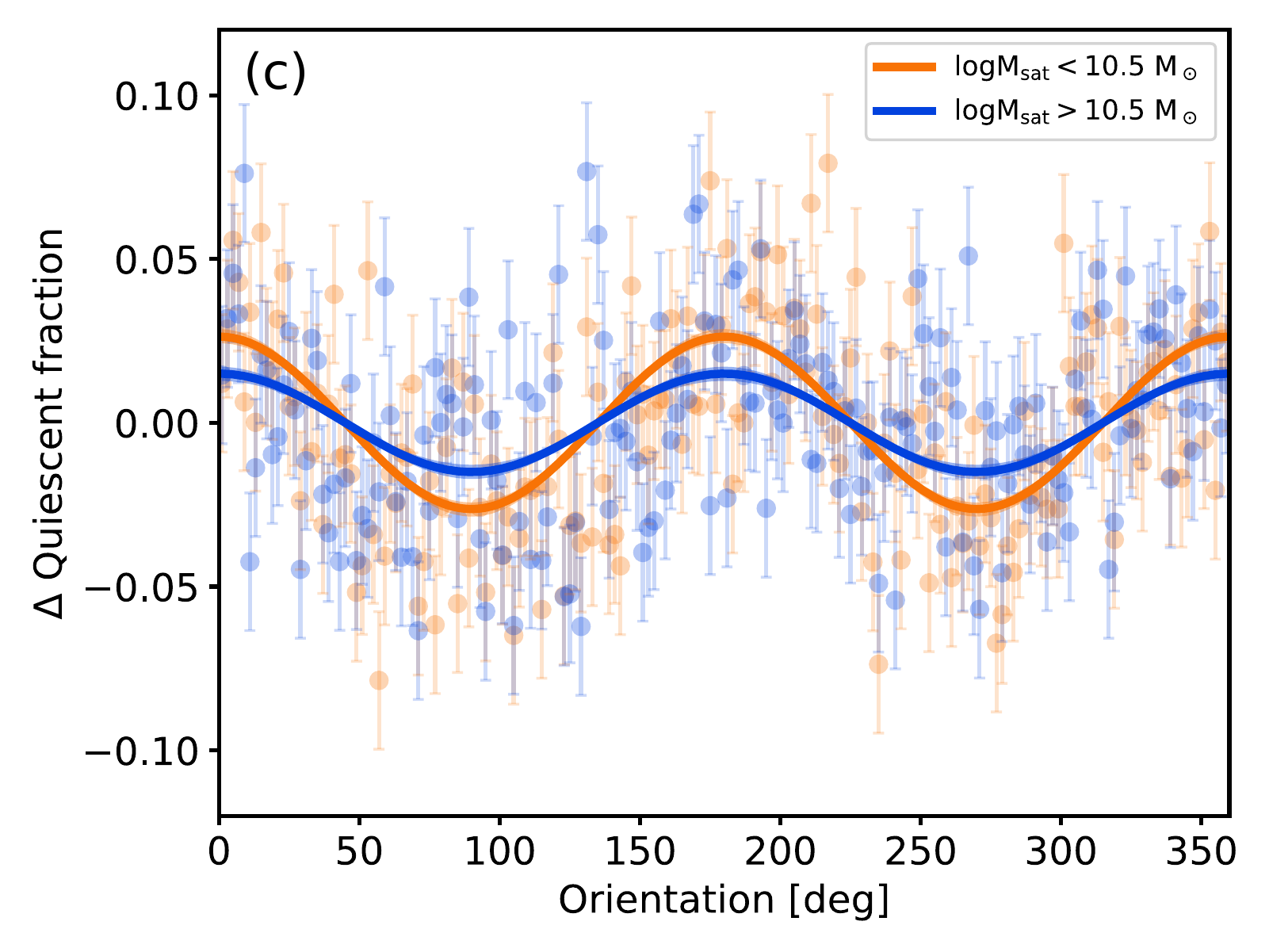}
    \includegraphics[height=4.4cm]{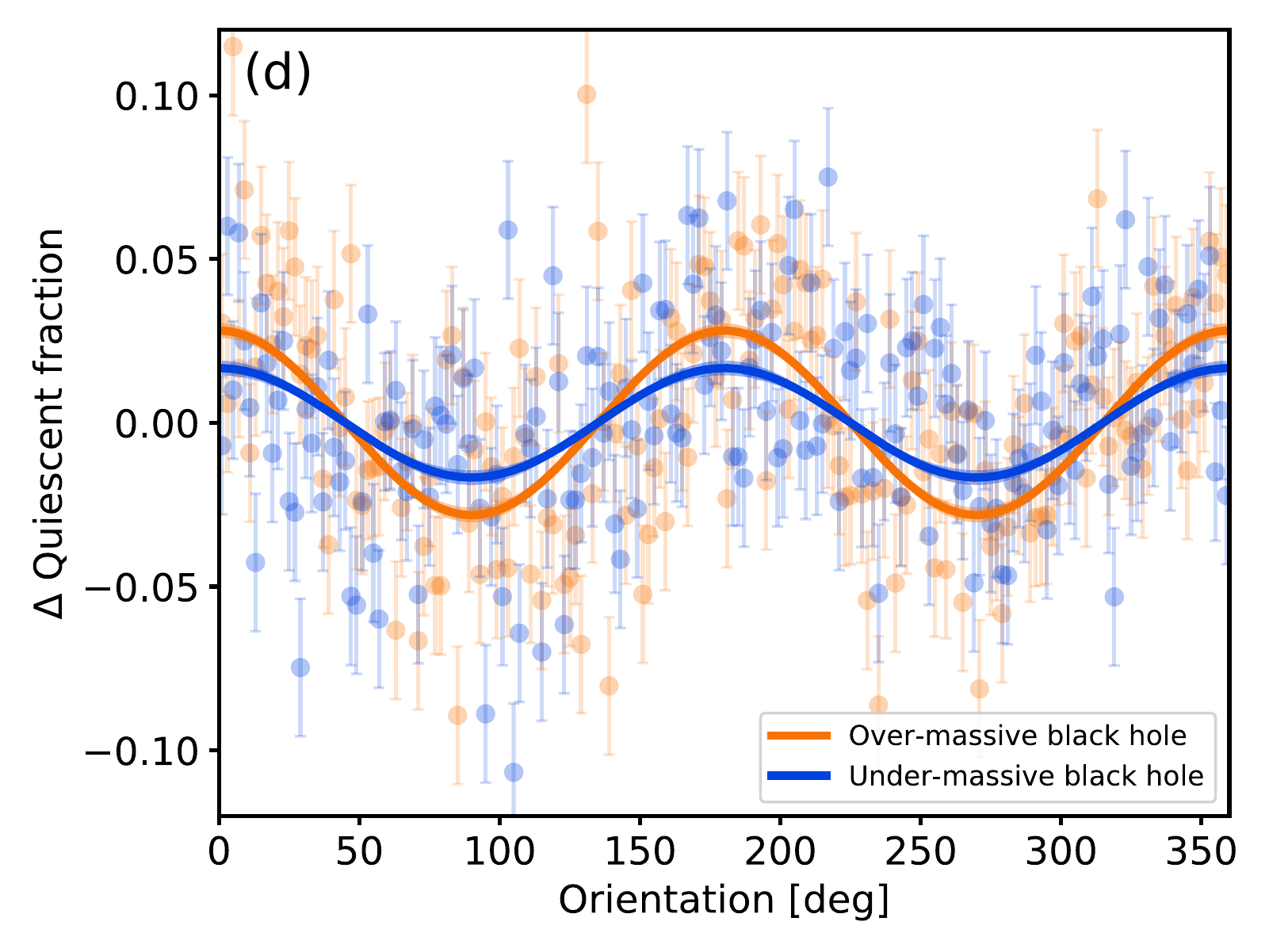}

    \includegraphics[height=4.4cm]{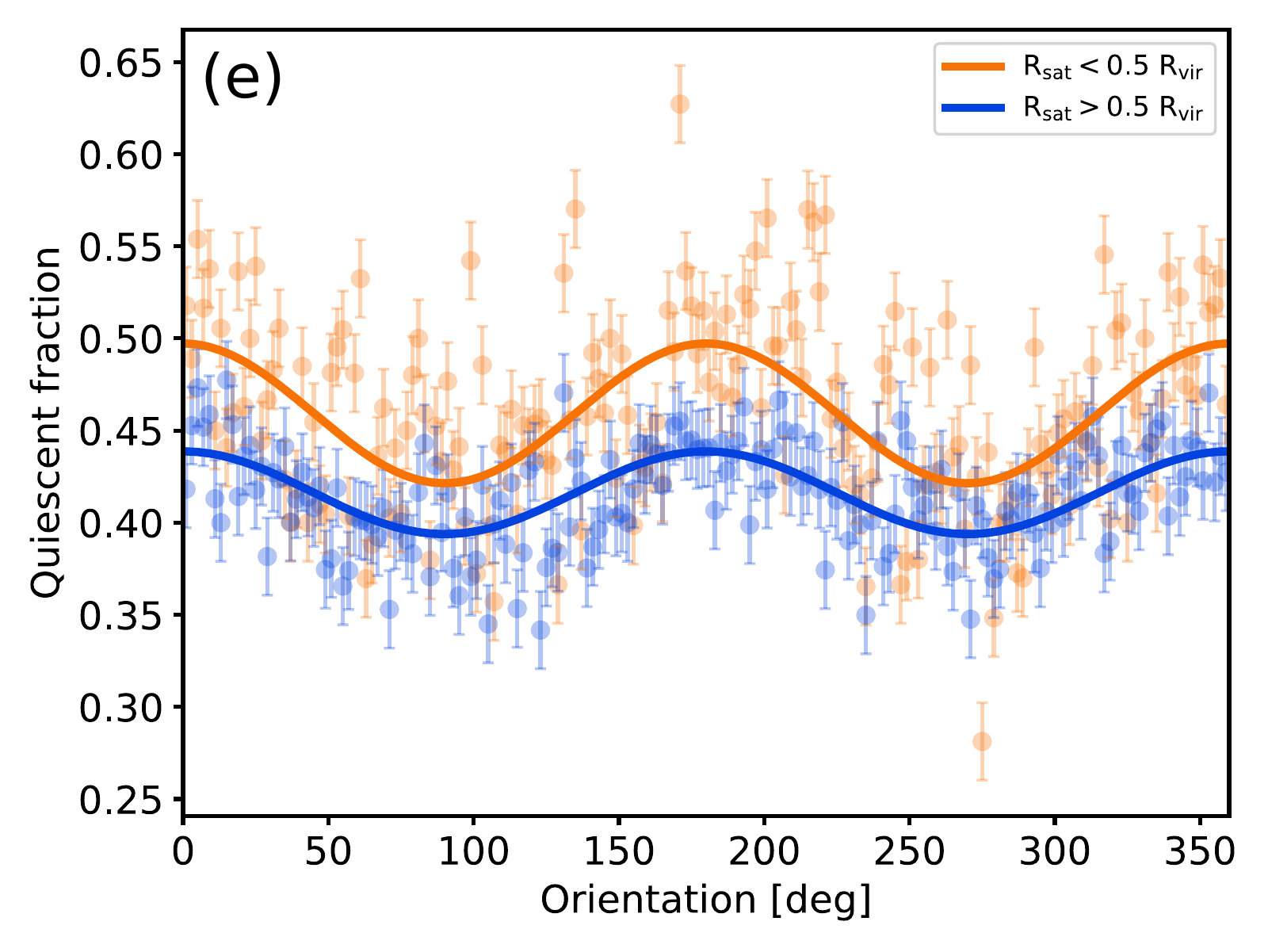}
    \includegraphics[height=4.4cm]{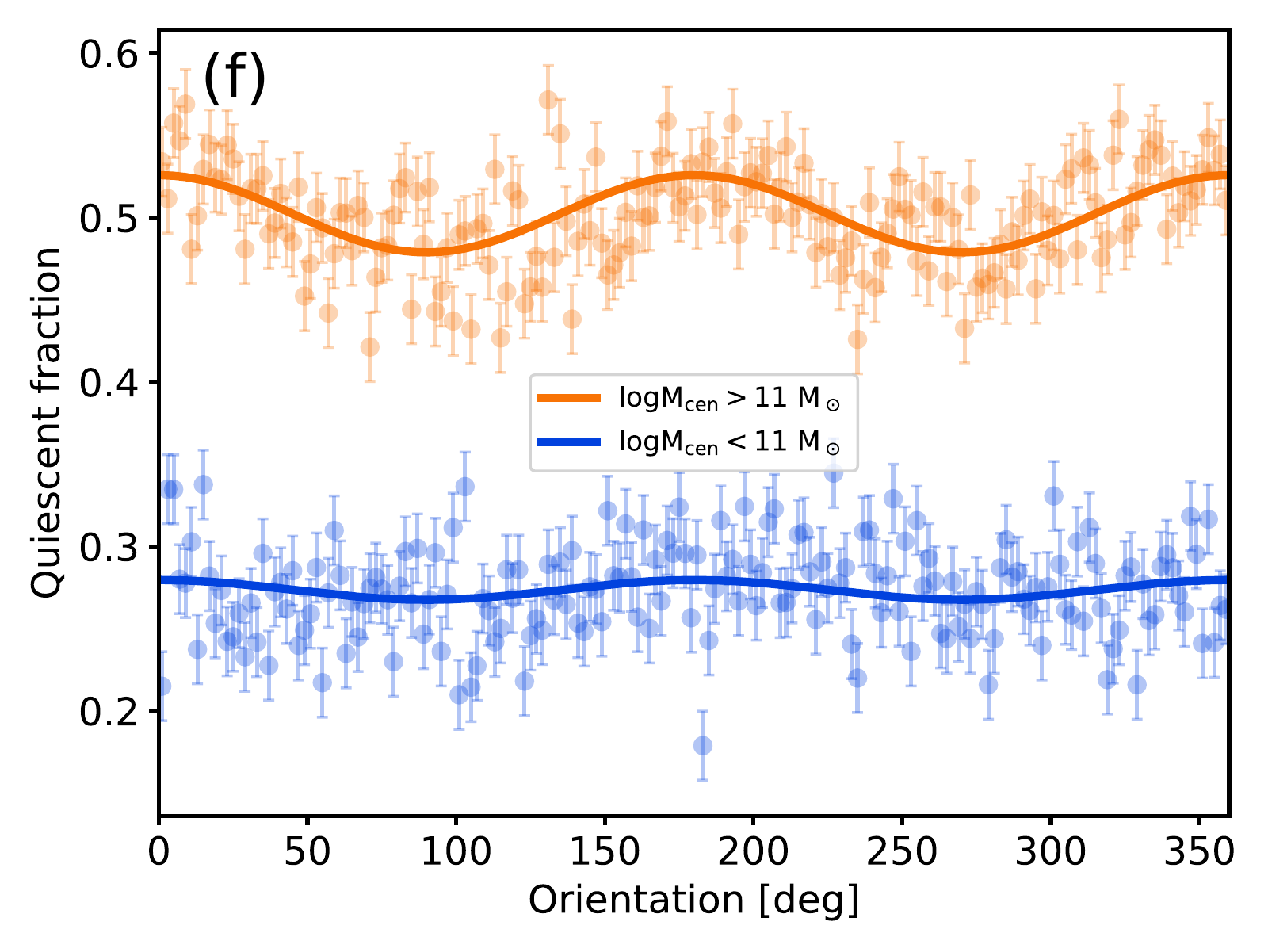}
    \includegraphics[height=4.4cm]{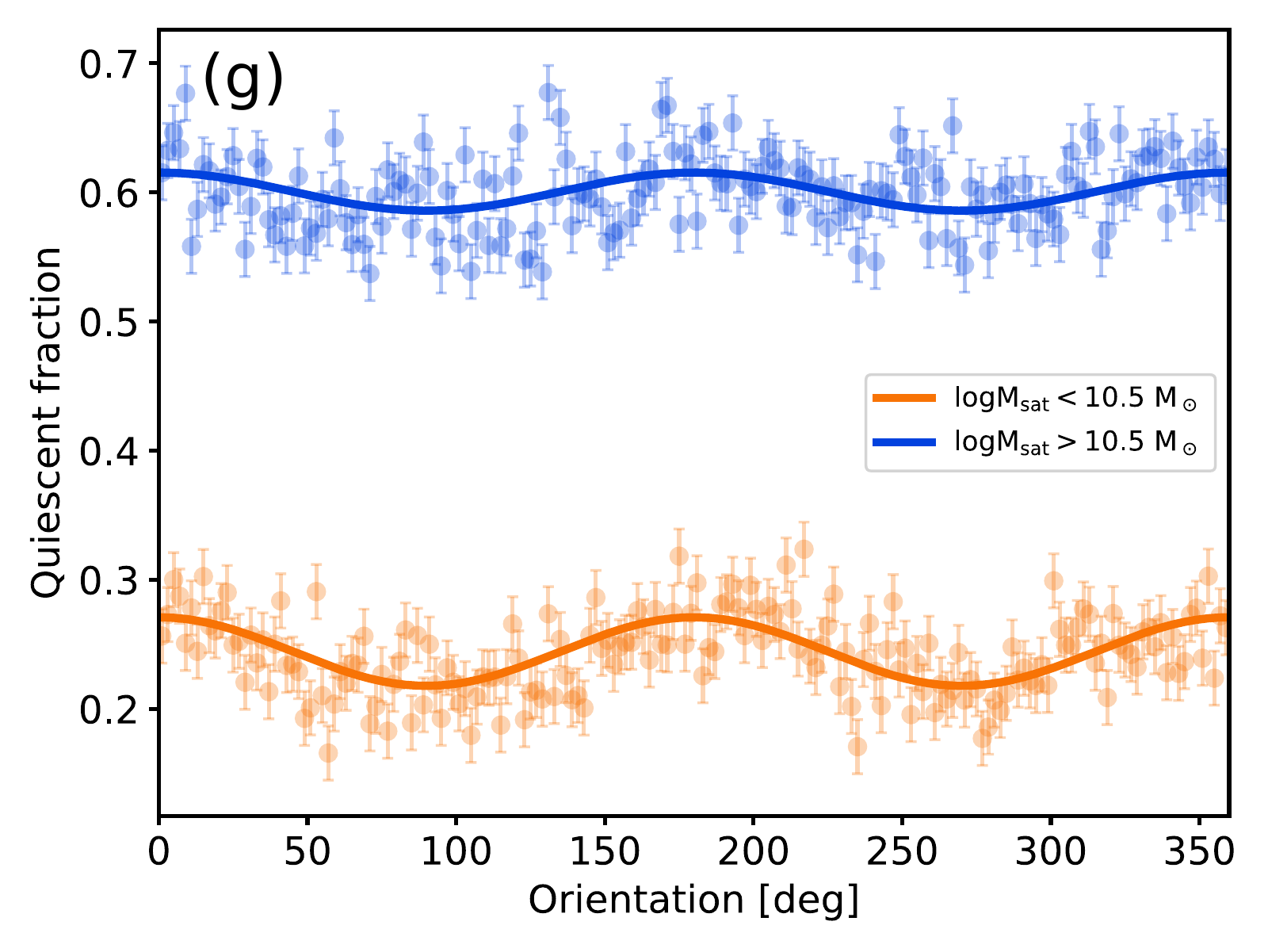}
    \includegraphics[height=4.4cm]{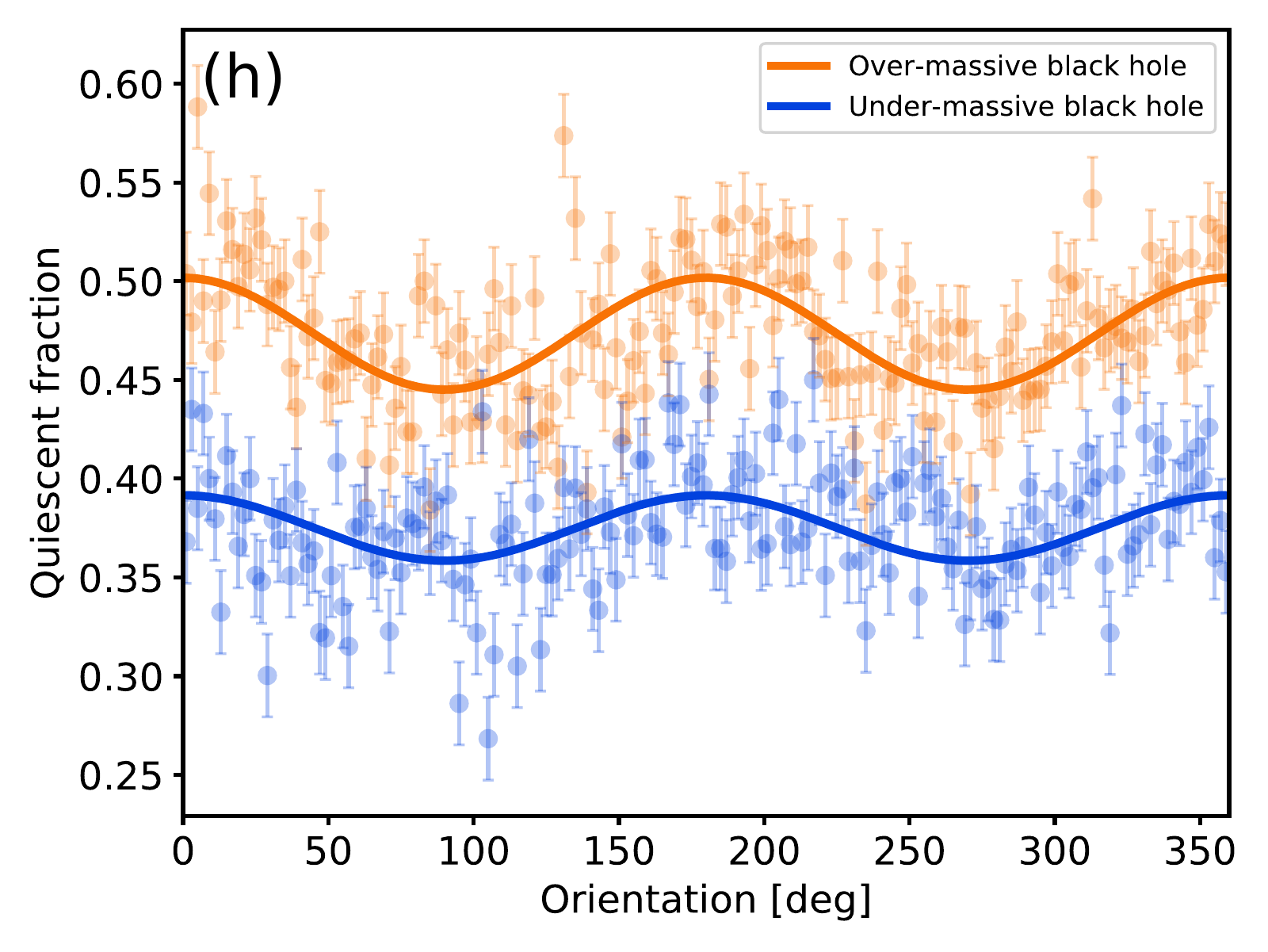}

    \end{center}
    \caption{{\bf Characterization of the SDSS signal.} In panel (a) we show that the modulation in the observed signal is higher for satellites closer to the centre (R$_\mathrm{sat} < 0.5$R$_\mathrm{vir}$, orange symbols) than for those satellites in the outskirts (R$_\mathrm{sat} < 0.5$R$_\mathrm{vir}$, blue symbols). In panel (b) the signal is stronger for halos with more massive centrals ($\log$M$_\mathrm{cen} > 11$\msun, orange symbols) compared to the signal observed in halos with less massive centrals ($\log$M$_\mathrm{cen} < 11$\msun, blue symbols).  In panel (c) less massive satellites  ($\log$M$_\mathrm{sat} < 10.5$\msun, orange symbols) exhibit a larger variation than more massive ones ($\log$M$_\mathrm{sat} > 10.5$\msun, blue symbols). Panel (d) reveals how the signal is also stronger in halos hosting more massive black holes in their center (orange symbols), compared to those with  relatively less-massive central black holes (blue symbols) Panels panels (e), (f), (g), and (h) are equivalent to (a), (b), (c), and (d) but without removing the offset between the different sub-samples.
    }
    \label{fig_methods:1}
\end{extfig*}

In addition, we also look at how the amplitude of the signal changes with the mass of the central galaxy, illustrated in Extended Data Fig.~\ref{fig_methods:1} panel (b). It is clear from this panel that the signal is stronger for satellites orbiting more massive centrals ($\log$ M$_\mathrm{cen} > 11$ \msun), with an amplitude of 0.024 $\pm 0.002$ compared to what is observed for halos with a lower mass central ($\log$ M$_\mathrm{cen} < 11$ \msun), where the amplitude is 0.007 $\pm 0.002$. The opposite behavior is however observed when splitting our sample according to the mass of the satellites, as demonstrated in panel (c). Interestingly, the modulation of the signal is stronger for lower-mass satellites ($\log$ M$_\mathrm{sat} < 10.5$ \msun) with an amplitude of 0.027 $\pm 0.002$. The amplitude measured for higher-mass satellites ($\log$ M$_\mathrm{sat} > 10.5$ \msun) is  0.014 $\pm 0.002$.

Here we also demonstrate what mentioned in the main text, namely that the strength of the observed modulation correlates with the (relative) mass of the super-massive black hole hosted by the central galaxy: see panel (d) of Extended Data Fig.~\ref{fig_methods:1}. Following previous works\cite{MN20}, black hole masses are estimated using the empirical \mbh--$\sigma$ relation \cite{vdb16}, and the stellar velocity dispersion of each central $\sigma$ is measured using the available SDSS spectroscopic data. At fixed halo mass, the amplitude of the observed signal is higher for those halos with a over-massive black hole in their centers (0.028 $\pm 0.002$). For halos hosting under-massive black holes, the observed amplitude in the signal is 0.016 $\pm 0.002$. The expected uncertainty in these black hole mass estimates is of $\sim0.3$ dex due the intrinsic scatter in the \mbh-$\sigma$ relation\cite{MN19}. In individual galaxies, this over-massive vs under-massive black hole metric has been now widely used to probe the interplay between black hole activity and star formation \cite{MN16,MN18b,MN20,Terrazas16,Terrazas17,Terrazas19,Dullo20,Li20}, further supporting a black hole-related origin for the observed signal. For completeness, panels (e), (f), (g), and (h) in Extended Data Fig.~\ref{fig_methods:1} show the same analysis as in panels (a), (b), (c), and (d) but without removing the offset between the different sub-samples.

\subsection{Additional metrics for the satellite star formation properties}

Although we have mostly focused on the fraction of quiescent galaxies as a proxy for the properties of the satellite populations, alternatives measurements can be explored that could provide further insights on the origin of the signal. For example, the mean specific star formation rate at each orientation, or the average distance of satellites with respect to the star formation main sequence provide a less {\it bimodal} and more continuous characterization of the satellite population. The behavior of these two quantities is shown in Extended Data Fig.~\ref{fig_methods:distance}, and also exhibits the characteristic modulation reported for the fraction of quiescent galaxies. The consistency between this figure and the trend shown in Fig.~\ref{fig:2} is an additional proof of the robustness of our results.

\begin{extfig} 
    \begin{center}
    \includegraphics[height=6.5cm]{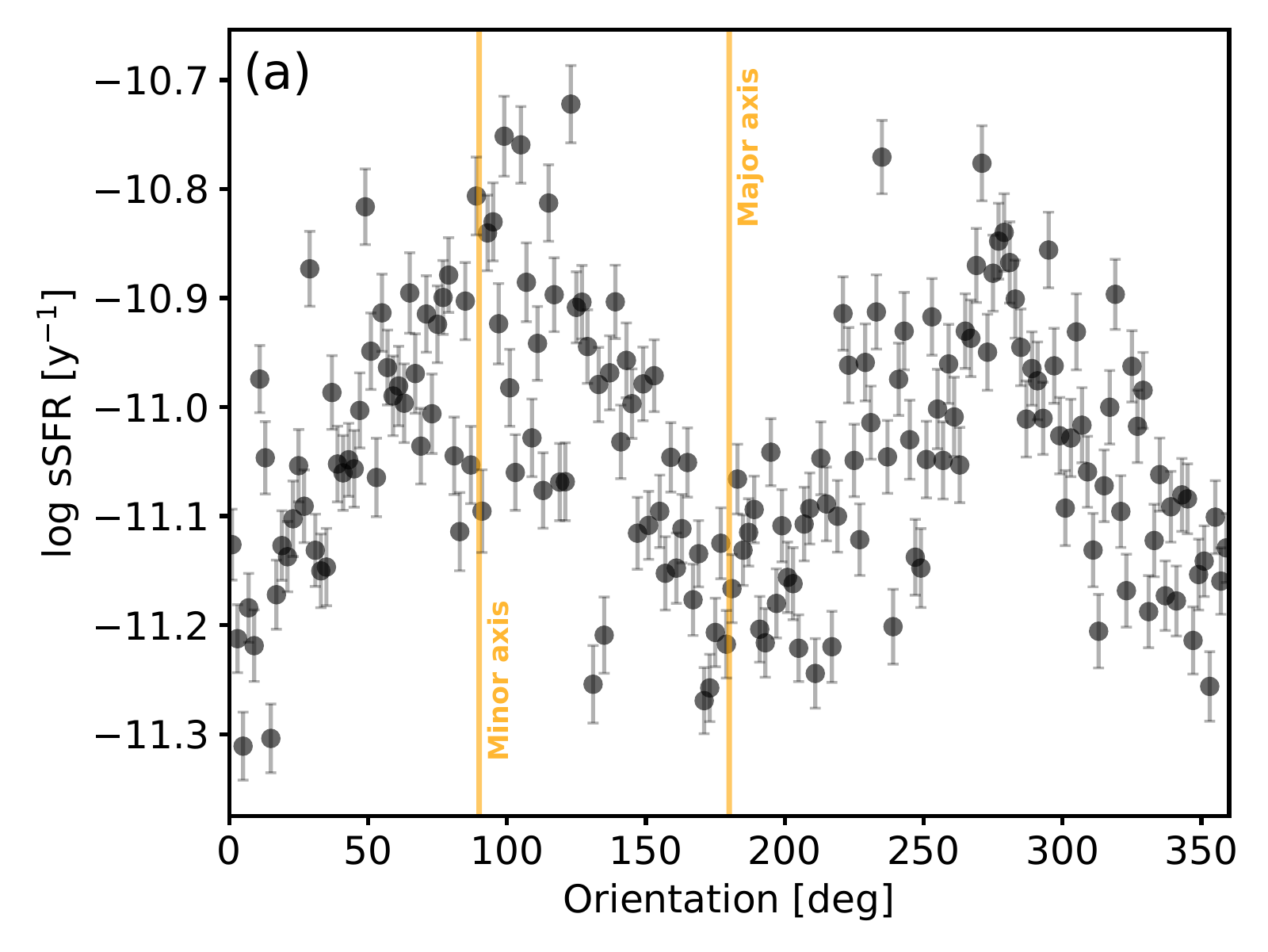}
    \includegraphics[height=6.5cm]{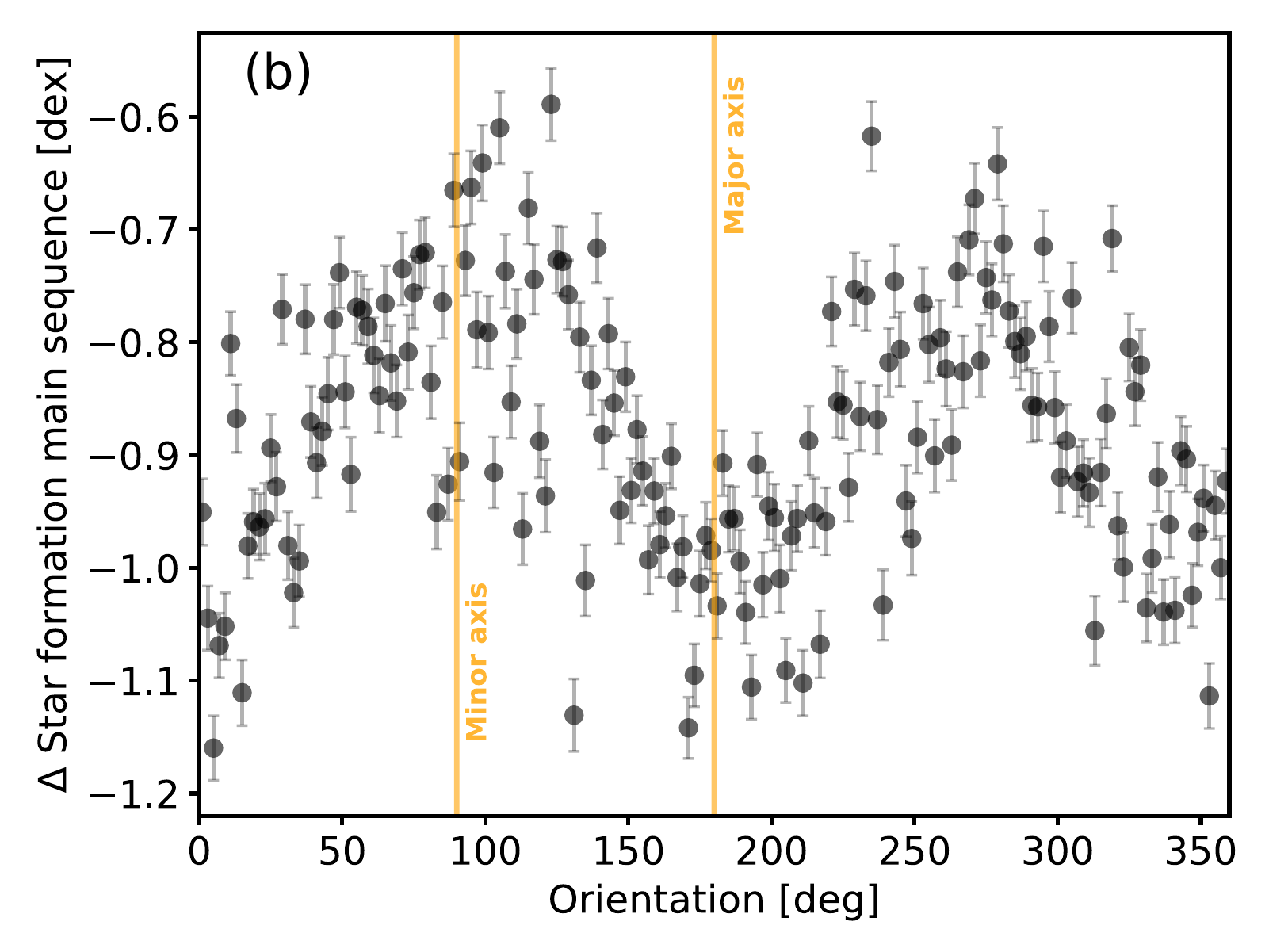}
    \end{center}
    \caption{{\bf Alternative metrics  for the characterization of SDSS satellites' star formation status}. The modulation observed in the average specific star formation rate (a) and distance from the star formation main sequence (b) of SDSS satellites closely follows that shown by the fraction of quiescent satellites in Fig.~\ref{fig:2}. Regardless of the metric used to characterize the star formation properties of satellite galaxies, there is a clear dependence on the orientation with respect to the central galaxy. Error bars indicate the 1$\sigma$ uncertainty and yellow lines mark the location of the minor and major axes.}
    \label{fig_methods:distance}
\end{extfig}

\subsection{Residual dependencies}

We have shown how the observed modulation in the fraction of quiescent satellites is stable and well-behaved. However, its peak-to-peak variation is only of $\sim5$\%, smaller than the expected change in the quiescent fraction when, for example, varying halo mass or cluster-centric distance. Thus, we also explored if the modulation could arise from variations in these properties. This was done by, first, characterizing the dependence of the quiescent fraction on M$_\mathrm{halo}$ and R$_\mathrm{vir}$ with a quadratic polynomial fit. Then, we use these fits to estimate the amplitude of the signal that would result from a variation in halo mass or cluster-centric distance with orientation. The result of these tests is shown in Extended Data Fig.~\ref{fig_methods:extra}.

\begin{extfig} 
    \begin{center}
    \includegraphics[height=6.5cm]{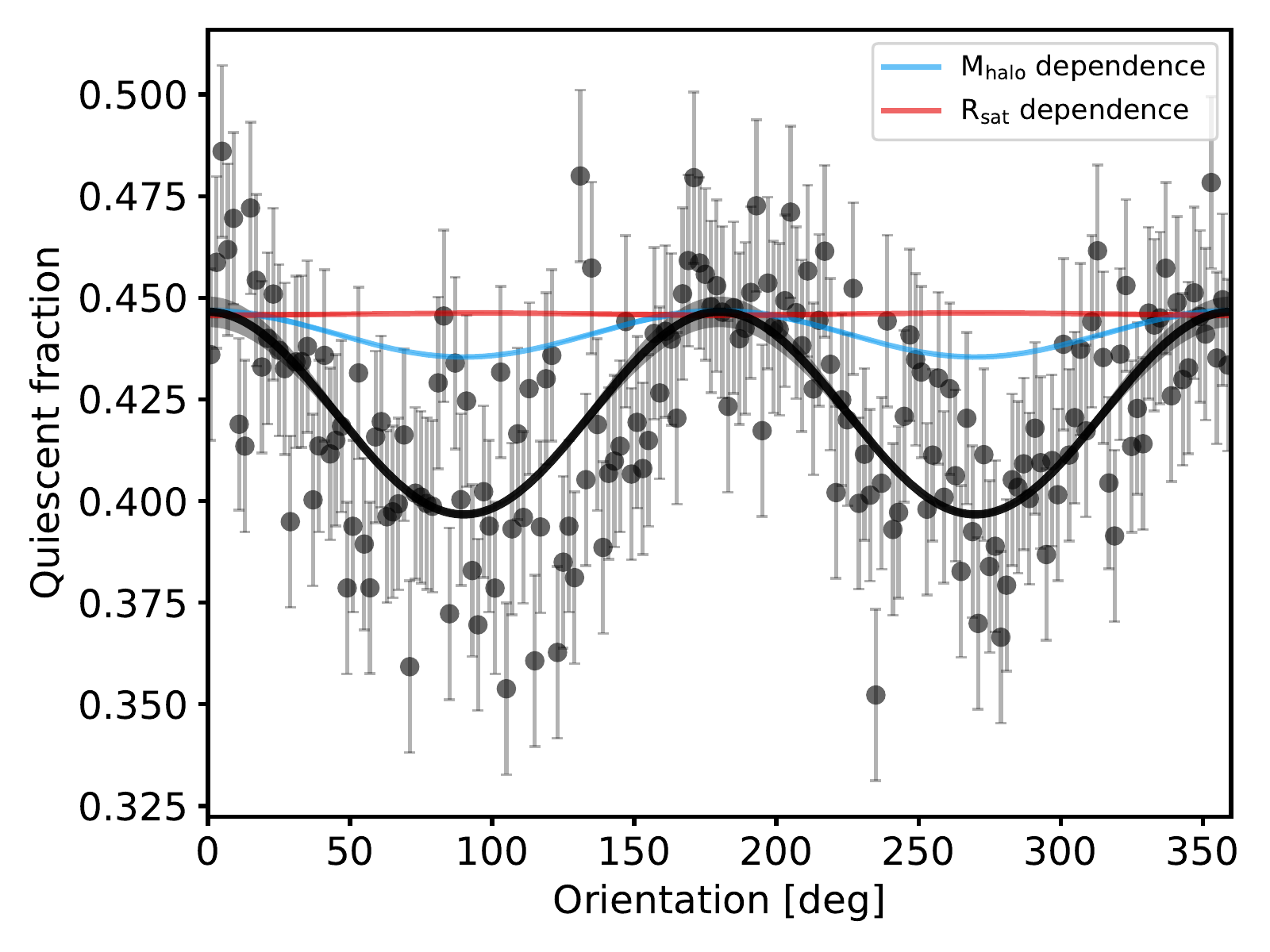}
    \end{center}
    \caption{{\bf Additional trends with halo mass and distance in SDSS.} As in Fig.~\ref{fig:2}, black symbols represent the observed modulation on the SDSS data. The blue line indicates the change in the quiescent fraction that could be expected because of the average halo mass dependence on orientation, which is significantly smaller than the reported one. Similarly, satellites along the minor axis are marginally closer to the central galaxy than along the major axis, leading to a negative and even weaker modulation, as shown by the red line.}
    \label{fig_methods:extra}
\end{extfig}

For both halo mass and average radial distance we find subtle trends with orientation, but at a level much lower than it would be required to explain the modulation shown in Fig.~\ref{fig:2}. In particular, the average halo mass along the minor axis tends to be slightly higher (0.05 dex) than along the major axis. Since more massive halos tend to host more massive galaxies, the observed variation in halo mass which would lead to a variation in the fraction of quiescent satellites of 0.005$\pm0.0001$, much smaller than the observed one. Once corrected for halo mass, no other secondary trends are found. Regarding the typical radial distance, satellites along the minor axis are marginally farther away from the central that along the major axis. In this case the variation is even weaker than for the halo mass, with a best-fitting amplitude of only $-0.0002\pm0.00005$. 

\subsection{Radial behavior}

Panel (b) in Fig.~\ref{fig:2} demonstrates that the observed signal is radially well-behaved, as it allows to measure the contours of iso-quiescent fractions. For simplicity, Fig.~\ref{fig:2} only shows one iso contour, but this idea can be further applied to different thresholds. In Extended Data Fig.~\ref{fig_methods:iso} we show these contours at three different levels f$_q$ = $\{ 0.36, 0.42, 0.48\}$

\begin{extfig} 
    \begin{center}
    \includegraphics[height=6.5cm]{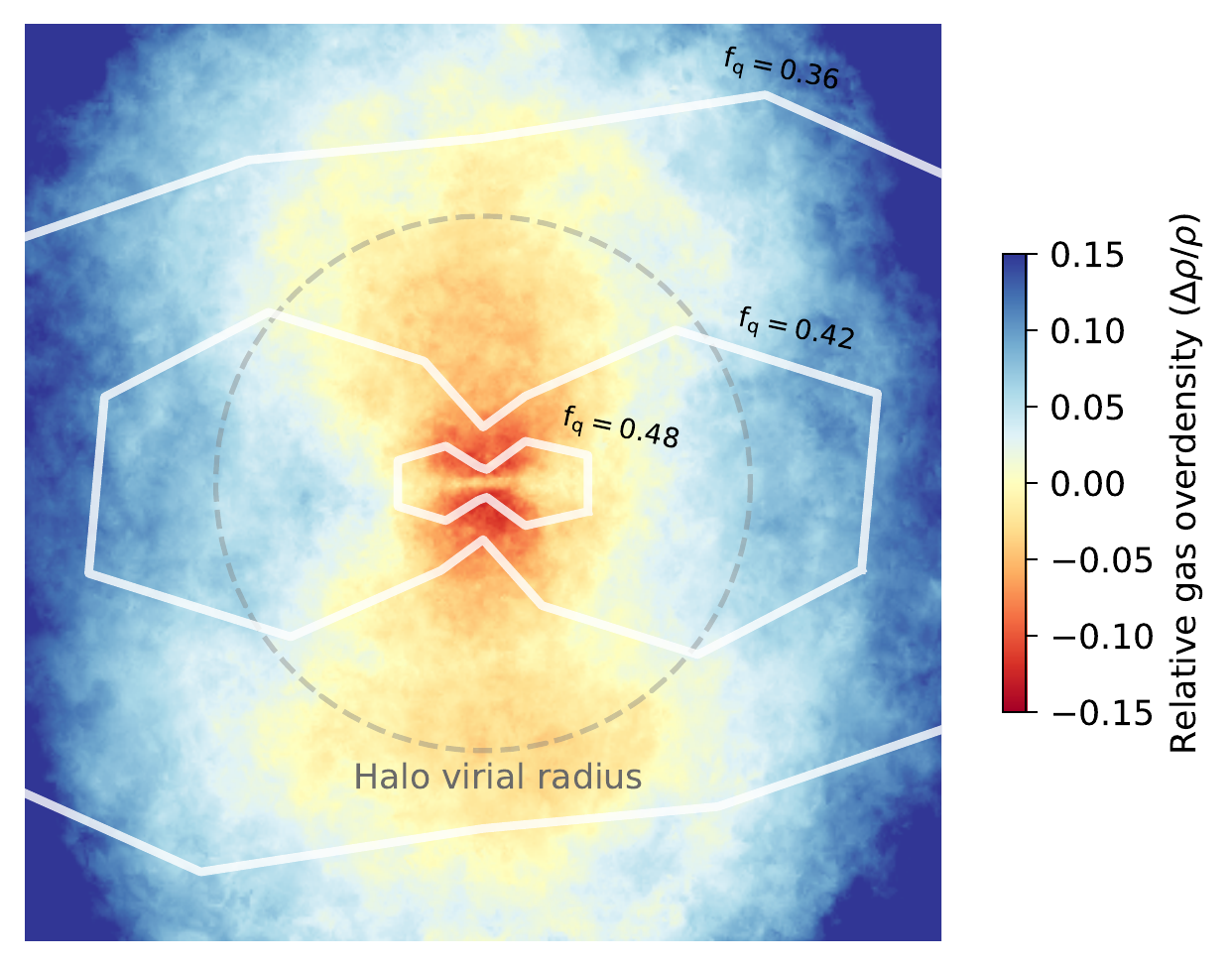}
    \end{center}
    \caption{{\bf Iso-quiescent fraction contours.} Similarly to Fig.~\ref{fig:4}, the contours of constant f$_q$ are shown, but this time at three different levels (f$_q$ = $\{ 0.36, 0.42, 0.48\}$). The background image corresponds to the IllustrisTNG gas over-density and the typical virial radius in the explored halo mass range is shown as a dashed grey circle.}
    \label{fig_methods:iso}
\end{extfig}

\subsection{Illustris vs IllustrisTNG simulation comparison} \label{methods:TNGvsIllustris}

As an additional way to investigate the origin of the observed signal, Extended Data Fig.~\ref{fig_methods:2} illustrates how the outcomes of the IllustrisTNG and Illustris cosmological numerical simulations compare. To marginalize over the possible differences in the overall quenching of galaxies between IllustrisTNG and Illustris\cite{Martina20}, also here we subtract the average quiescent fraction of both datasets.

As mentioned in the text, the  TNG100 run of the IllustrisTNG series and the Illustris simulations both sample an approximately similar cosmological volume of 100 comoving Mpc: they in fact evolve the same initial conditions -- but for small variations in the values of the adopted cosmological parameters -- and have been performed at very similar numerical resolution. Illustris and IllustrisTNG differ in certain aspects of the underlying galaxy formation model\cite{Pillepich18}.
In addition to an accurate treatment of the magneto-hydrodynamics within the simulation, IllustrisTNG improves upon the original Illustris galaxy formation model on three fronts (see Section 2.3 in \cite{Pillepich18}): chemical enrichment, galactic winds, and black hole feedback. The first one is unrelated to the topic of this work and the physical functioning of the stellar feedback under the form of galactic winds is similar albeit not identical in the two models. On the other hand, different mechanisms are invoked and implemented for the feedback from the super massive black holes in the two models, specifically at low accretion rates\cite{Weinberger17}. In particular, in both Illustris and IllustrisTNG, SMBH feedback is implemented by invoking three mechanisms\cite{Weinberger17, Pillepich18}: thermal energy injection at high mass accretion rates, mechanical feedback at low mass accretion rates, and a sort of radiative feedback where gas cooling is further modulated by the radiation field of nearby AGNs. The two simulation models differ substantially only in the way the mechanical, low-accretion rate channel functions\cite{Weinberger17}: in IllustrisTNG, kinetic energy is injected into the surrounding gas as a pulsated wind, oriented in a different random direction at each SMBH timestep, and thus isotropic when averaged across any cosmological time scale of relevance; in contrast, in Illustris, thermal energy is injected in a highly bursty fashion into $\sim50-100$ kpc bubbles of gas at distances of a few $10-100$s of kpc from the central galaxy. Such differences -- whereby, in one case, AGN feedback affects gas at large radii rather than acting directly from the innermost regions of galaxies as in the case of IllustrisTNG --, result in important differences not only in the onset of galaxy quenching in central galaxies\cite{Nelson18, martina19} but also in the properties of the circumgalactic medium (e.g. gas density)\cite{Genel14,Pillepich18,Kauffmann_2019,Yun19,Lim20}. This implies that, in practice, by comparing in Extended Data Fig.~\ref{fig_methods:2} Illustris and IllustrisTNG, we are effectively testing and varying the effects of different black hole feedback mechanisms on the satellite population. 
\begin{extfig} 
    \begin{center}
    \includegraphics[height=6.5cm]{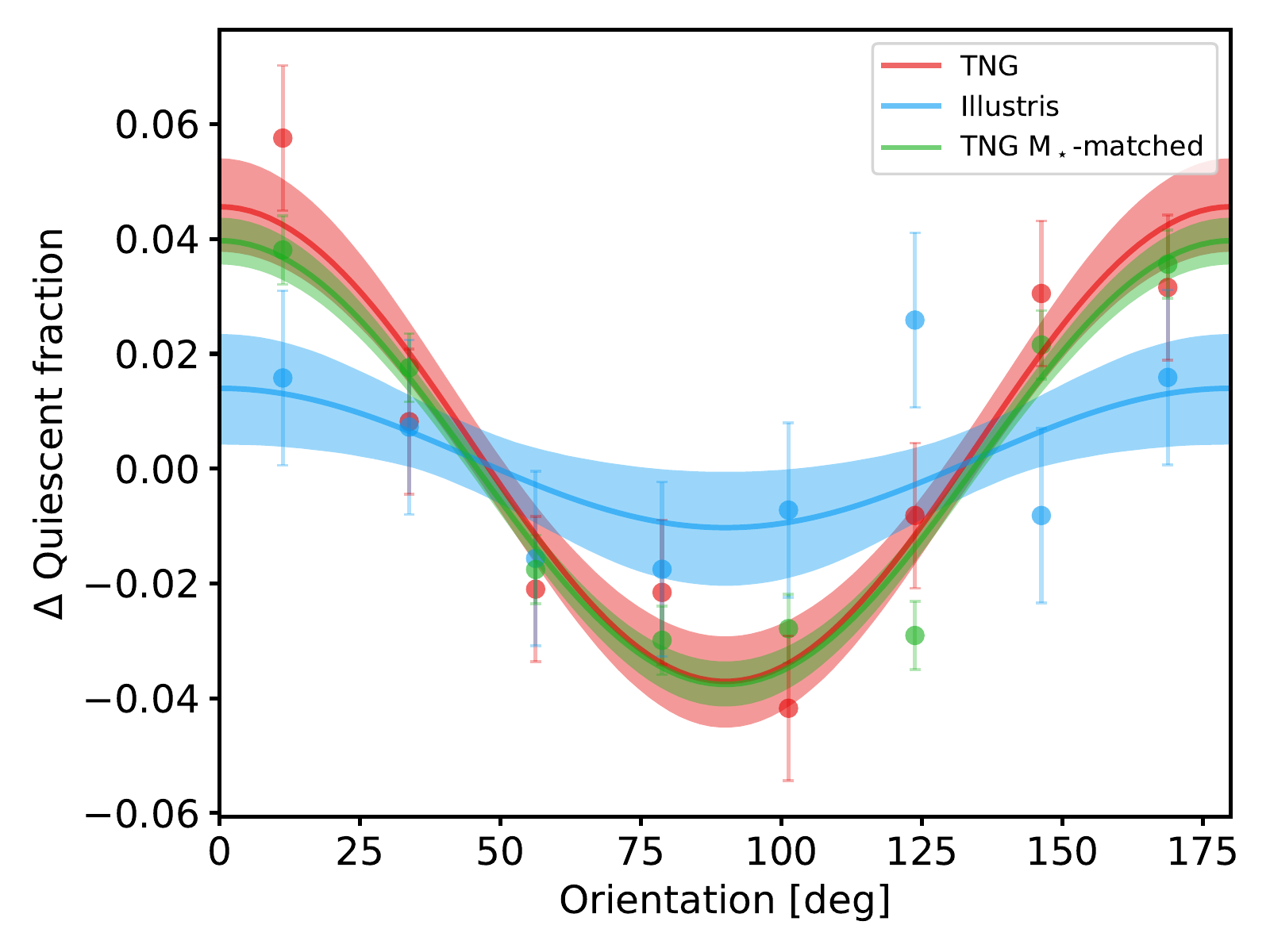}
    \end{center}
    \caption{{\bf IllustrisTNG vs Illustris comparison.} Modulation in the fraction of quiescent galaxies for the IllustrisTNG (namely, TNG100, red symbols) and the original Illustris (blue symbols) simulations. In green, the signal is shown for a sample of IllustrisTNG satellites with the same mass distribution as those in Illustris, to asses the possible effect of a mass bias between the two simulations. Both simulations probe a similar $\sim$ 100 Mpc comoving cosmological volume and thus share the same large-scale structure properties, while the treatment of black hole growth and feedback is the most relevant difference between the two. However, it is clear that the amplitude of the modulation is significantly higher in IllutrisTNG (0.032$\pm0.004$) than in Illustris (0.013$\pm0.007$).}
    \label{fig_methods:2}
\end{extfig}

For the blue curve in Extended Data Fig.~\ref{fig_methods:2}, Illustris galaxies are considered within similar ranges of host halo mass, and stellar mass of satellites and centrals as for the TNG100 sample (red curve). The signal is clear in IllustrisTNG, with an amplitude of 0.032$\pm0.004$, whereas in Illustris the signal is weaker (0.013$\pm0.007$), consistent with almost no variation. Note that, since the number of satellites in these large simulated volumes is still small compared to SDSS, we symmetrize the data along 180\textdegree and the binning is also coarser. As the assembly of the larger scale-structure is the same in the two simulations, this piece of evidence favors the proposed scenario whereby feedback from super-massive black holes is responsible for the anisotropic quenching signal, through its effects on the properties of the circumgalactic medium.

Note that since the mass function of Illustris differs from that of IllustrisTNG\cite{Pillepich18b}, in principle, it could be possible that the larger amplitude in IllustrisTNG is due to a bias in the mass distribution even if galaxies were selected to cover the same mass range. Green symbols in Extended Data Fig.~\ref{fig_methods:2} show the variation in the fraction of quiescent IllustrisTNG satellites, but intentionally selected to reproduce the stellar mass distribution of satellites in Illustris. The fact that the amplitude of this signal remains unchanged for this M$_\star$-matched sample suggests that a mass bias is not responsible for the differences between Illustris and IllustrisTNG.

\subsection{Ejective AGN feedback at the origin of the quenching directionality in IllustrisTNG} \label{methods:TNGteachings}

\begin{extfig} 
    \begin{center}
    \includegraphics[height=6.5cm]{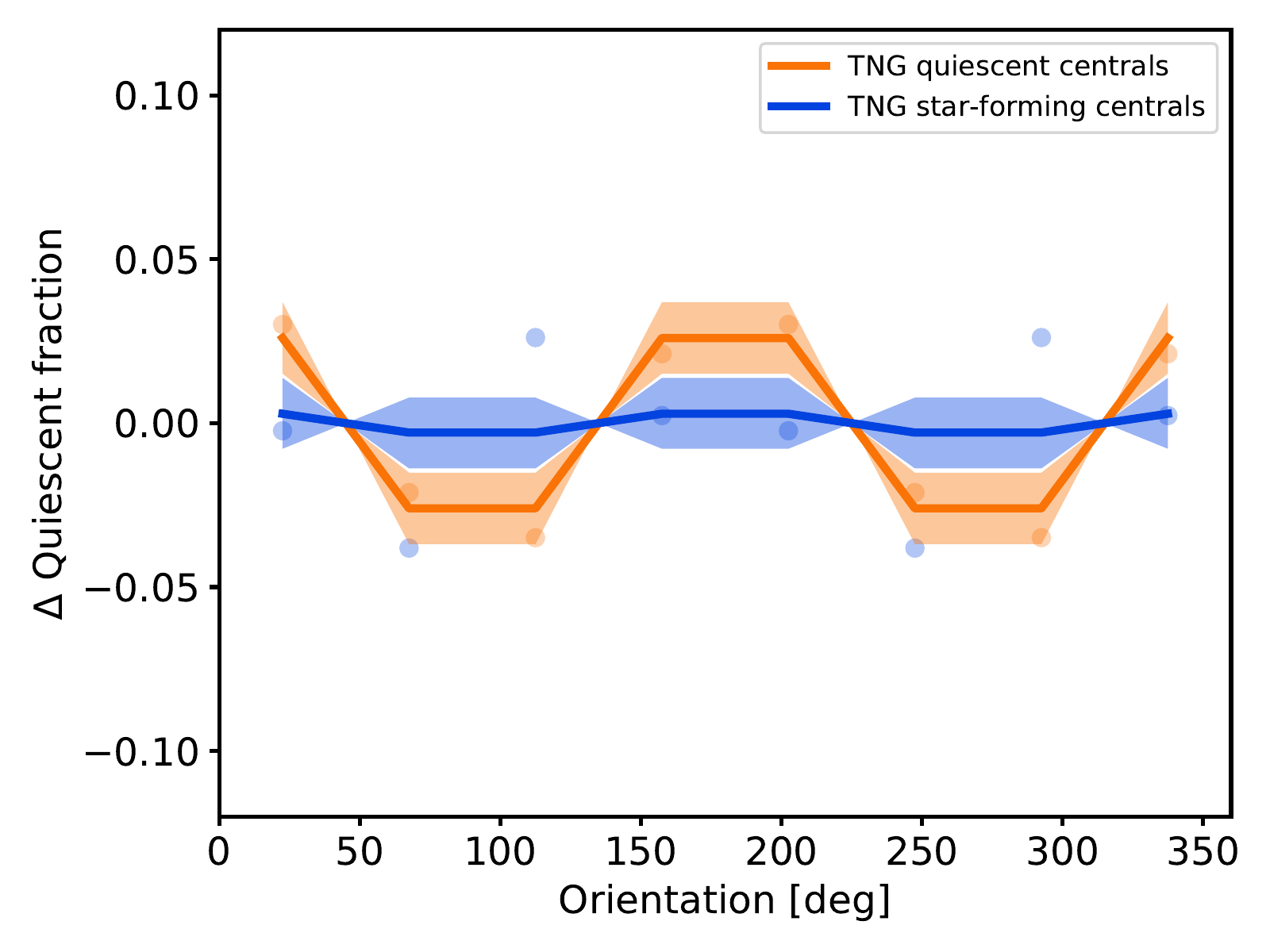}
     \includegraphics[height=6.5cm]{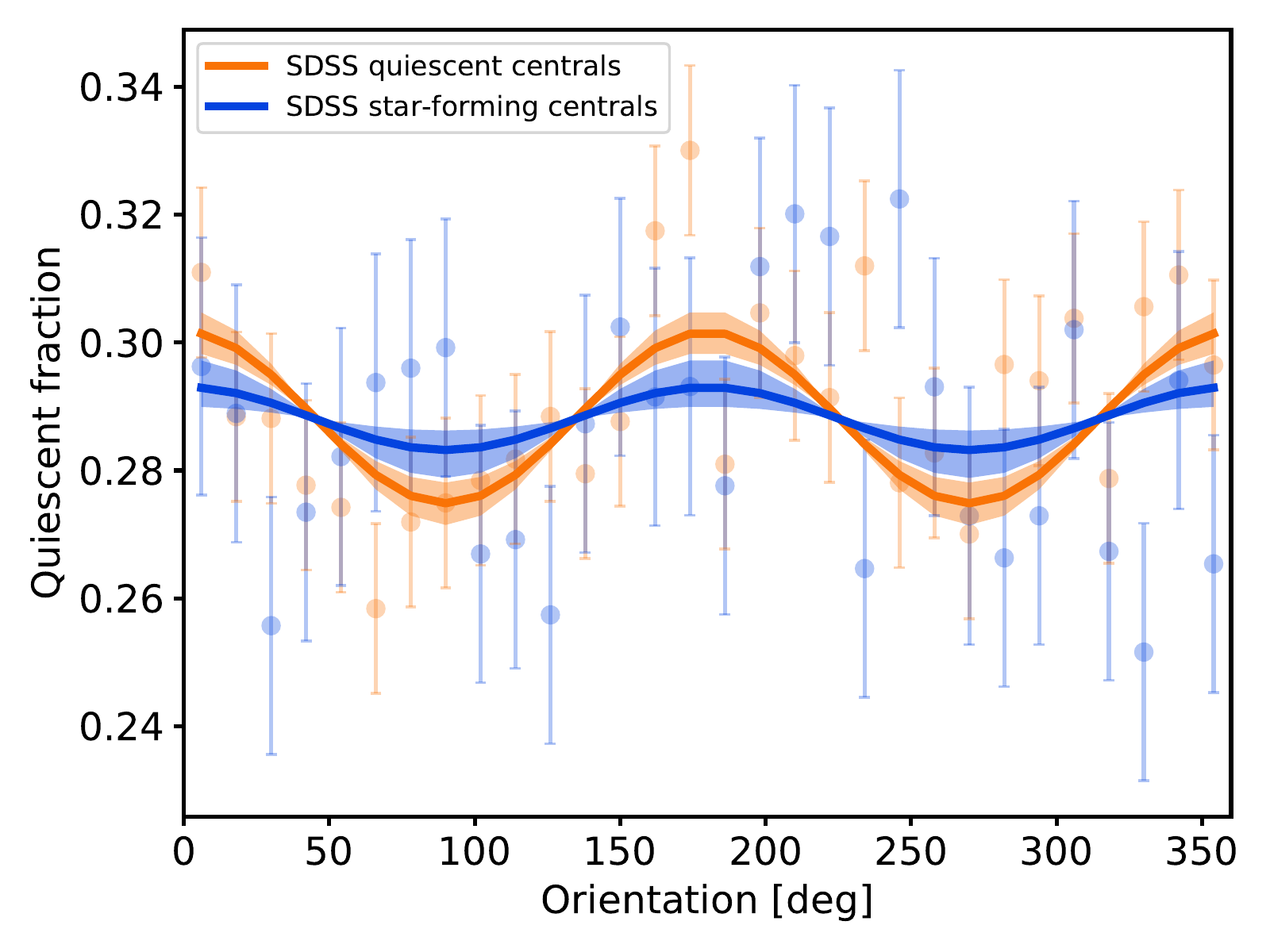}
    \end{center}
    \caption{{\bf Quiescent vs star-forming centrals in  IllustrisTNG (a) and SDSS (b).} In panel (a), at a fixed central stellar mass of $\sim \log$ M$_\mathrm{cen} = 10.5$ \msun, the modulation in the fraction of quiescent satellites in TNG100 is shown for star-forming (blue) and quiescent (orange) centrals. Although with a limited number of satellites, the modulation in the signal appears to be stronger for quiescent centrals than for star-forming ones. Since {\it quiescentness} in IllustrisTNG is a strong indication of an effective black hole feedback, the fact that the signal is stronger for quiescent galaxies is also an indication of the proposed AGN-related origin for the observed quenching directionality. The modulation in the fraction of quiescent satellites is shown for star-forming (blue) and quiescent (orange) centrals in panel (b) but this time for SDSS galaxies, again of $\log$ M$_\mathrm{cen} = 10.5$ \msun. The observed modulation is stronger for quiescent than for star-forming centrals as seen in IllustrisTNG. Solid lines and shaded areas indicate the best-fitting trends and 1-$\sigma$ confidence interval, respectively.}
    \label{fig_methods:3}
\end{extfig}

In IllustrisTNG, whether a massive galaxy is quenched or not is a direct indication of the effectiveness of black hole feedback at low accretion rates \cite{Nelson18}. Thus, a complementary way to asses the role of black hole feedback in shaping the observed modulation is by comparing the amplitude of the observed signal for quiescent and star forming centrals at fixed stellar mass. The outcome of this comparison is shown in Extended Data Fig.~\ref{fig_methods:3}, panel (a), for central galaxies with stellar masses of $\log$ M$_\mathrm{cen} = 10.5$ \msun. This particular mass range is selected so that both star-forming and quiescent centrals are abundant enough. Due to the additional constraints in stellar mass and SFR, the number of available satellites is rather small (typically $\sim100$ per angular bin) and therefore the angular binning is coarser. 

It is apparent from Extended Data Fig.~\ref{fig_methods:3} that a systematic difference exists between the amplitude of the modulation in IllustrisTNG depending on the SFR of the central galaxy, as the signal is stronger for groups and clusters with a quiescent central galaxy, vanishing for satellites hosted by star-forming centrals. Since {\it quiescence} in massive IllustrisTNG galaxies is the consequence of an efficient and long-lasting AGN activity, the fact that the modulation is stronger for quiescent centrals further supports the black hole-related origin of the observed signal. Because of the low number of satellites in this test, however, the implications of Extended Data Fig.~\ref{fig_methods:3} should not be overstated.

For completeness, we also investigate if the differences between quiescent and star-forming centrals shown in Extended Data Fig.~\ref{fig_methods:3}, (a), is present in the SDSS observed data. As proved by Extended Data Fig.~\ref{fig_methods:3}, panel (b), this in fact the case for SDSS central galaxies with stellar masses of $\log$ M$_\mathrm{cen} = 10.5$ \msun.

In IllustrisTNG, the nature of the observed signal can be also tested by looking at how it depends on the amount of energy radiated, and on the mode in which this energy was radiated. In Extended Data Fig.~\ref{fig_methods:energy}, we select for TNG100 galaxies around $\log$ M$_\mathrm{cen} = 10.5$ \msun and we split them in two groups, depending on the cumulative energy injected by their super-massive black holes. Panel (a) shows how the modulation in the quenching directionality signal is stronger for satellites around central galaxies whose black holes have injected, at a given stellar mass, more total (kinetic plus thermal) energy than the average: these results suggested by IllustrisTNG further link the observed signal with the black hole activity. Panel (b) shows the result of splitting central galaxies but only according to the kinetic energy injected by their black holes when accreting at low rates. In this case, the signal almost vanishes for centrals that have undergone a relatively low amount of kinetic energy, once again supporting the idea that the signal is due to the ejective nature of black hole feedback, at least in the IllustrisTNG model. Whether, on the other hand, implementations of thermal and radiation SMBH feedback that are different from those adopted within Illustris and IllustrisTNG can reproduce the modulation of the fraction of quiescent satellites observed in SDSS remains to be determined.

Complementary, panel (c) in Extended Data Fig.~\ref{fig_methods:energy} shows the result of splitting IllustrisTNG galaxies by their relative black hole mass. As for the SDSS data (see panel (d) in Extended Data Fig.~\ref{fig_methods:1}), the signal is stronger for satellites orbiting over-massive black holes centrals than for under-massive black hole ones, reinforcing the proposed connection between black hole activity and the observed modulation in the fraction of quiescent galaxies.

In the observed SDSS data, the orientation angle between central and satellite galaxy is a projected one. That is, two satellites at apparently the same orientation may be actually located at a different 3D angle with respect to the central galaxy. Thus, the true underlying modulation is expected to be higher than the observed (projected) one. In order to explore this effect, we make use of the three-dimensional information provided by the IllutrisTNG simulation to de-project the observed modulation in the fraction of quiescent satellites. The result of this de-projection is shown in panel (d) of Extended Data Fig.~\ref{fig_methods:energy}. As expected from a three-dimensional effect such as the proposed black hole feedback mechanism, the amplitude of the signal, once de-projected, is larger than in the projected space.

\begin{extfig*} 
    \begin{center}
    \includegraphics[height=6.5cm]{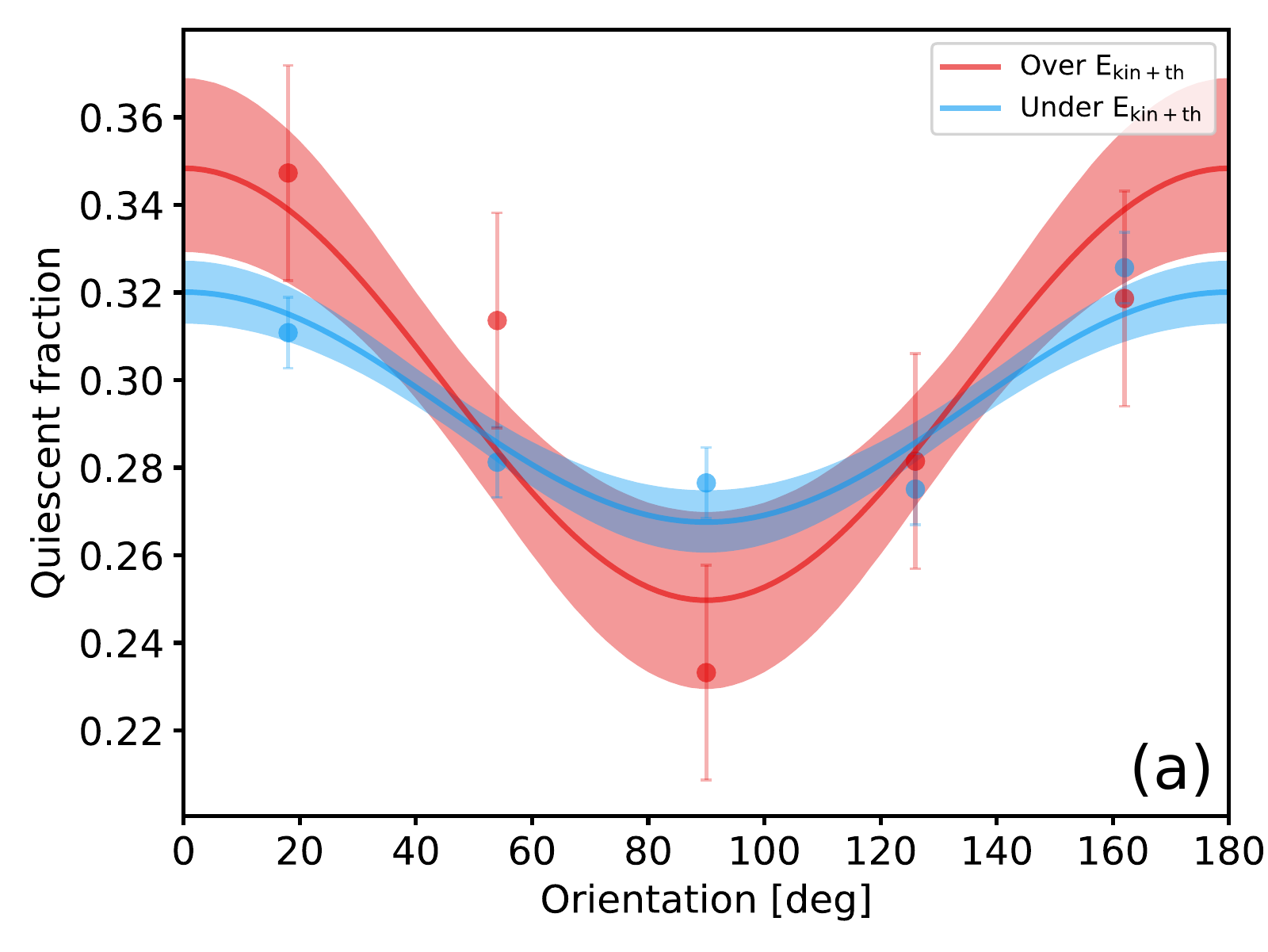}
    \includegraphics[height=6.5cm]{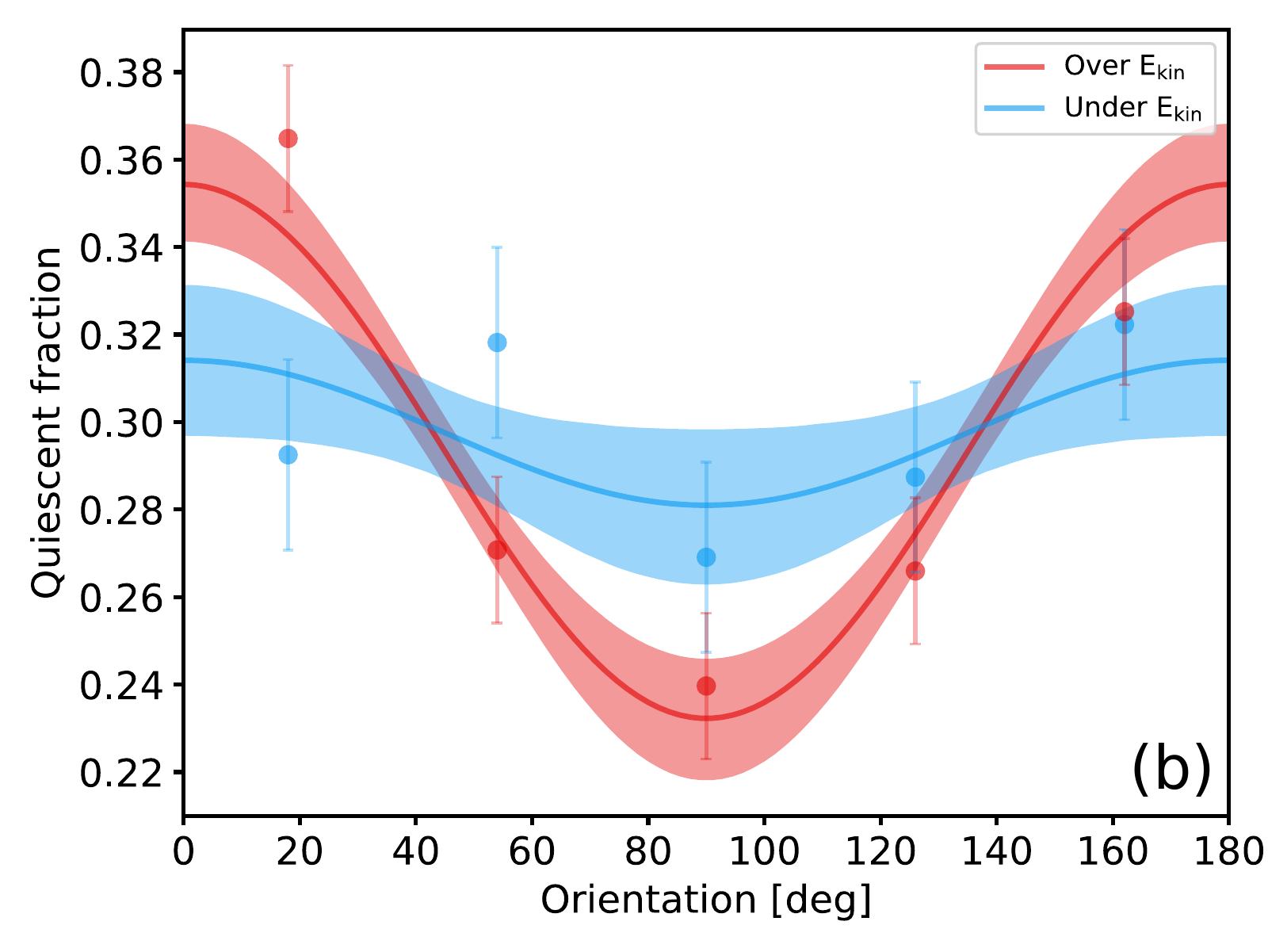}
    \includegraphics[height=6.5cm]{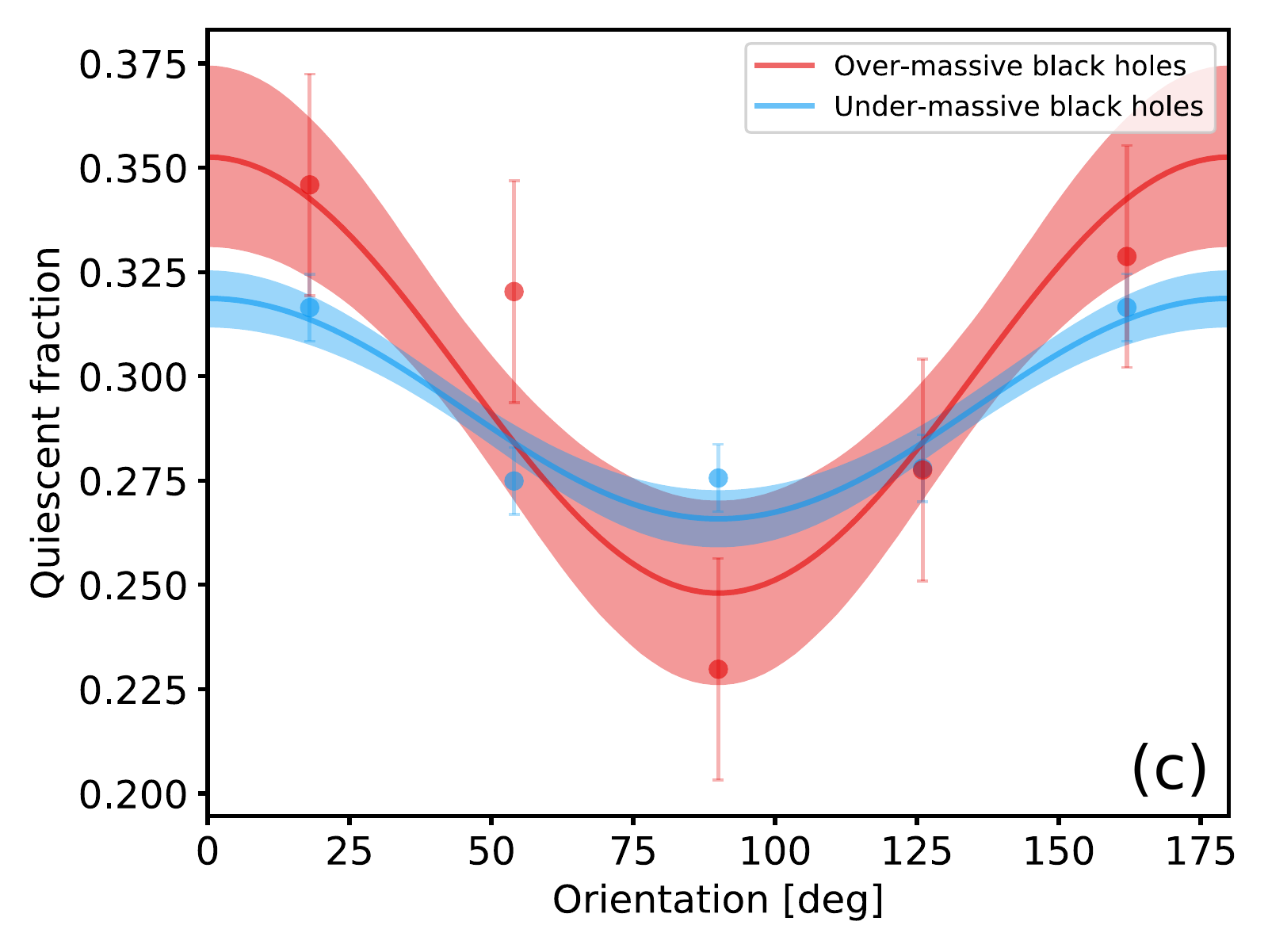}
    \includegraphics[height=6.5cm]{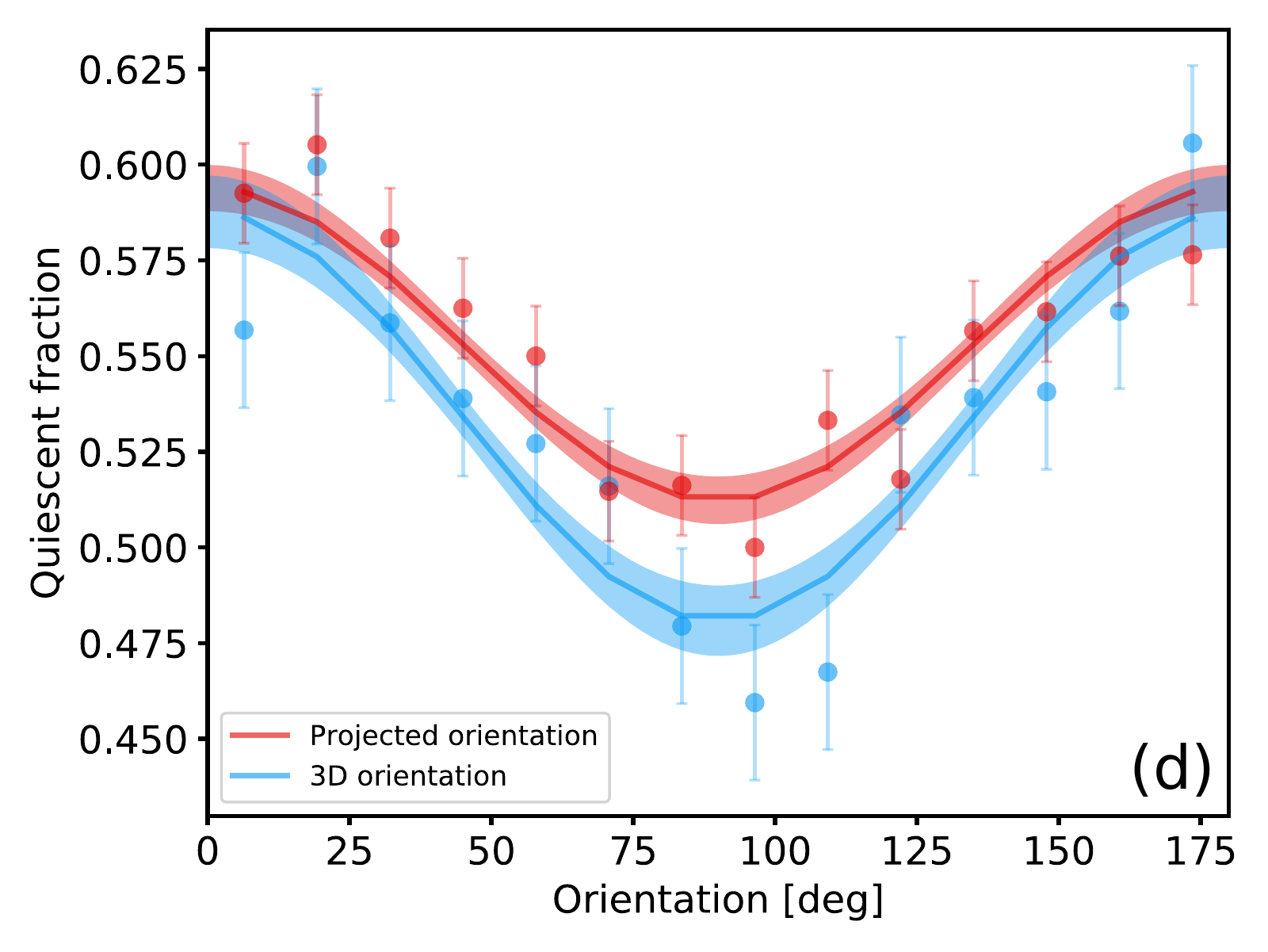}
    \end{center}
    \caption{{\bf Dependence of the signal on the energy emitted by the super-massive black holes in IllustrisTNG.} Panel (a) shows the fraction of quiescent satellites around centrals whose black holes have injected, relatively to their mass, more (red) and less (blue) total energy. Similarly, Panel (b) shows the same separation but in this case considering only the {\it kinetic} energy injected by the black holes. In both cases, the amplitude of the modulation is stronger when the total (panel a) and kinetic (panel b) energy released by the central black holes increase. Similar to Extended Data Fig.~\ref{fig_methods:1}, panel (c) shows how the signal in IllustrisTNG depends on the relative mass of the central black hole, being stronger for more over-massive black hole galaxies. Finall, panel (d) shows, in red, the observed signal in IllustrisTNG, and in blue the de-projected using the underlying 3D satellite distribution. Note that in this panel we did not impose any cut in central stellar mass and therefore absolute values are different from the other panels. Error bars and shaded areas represent 1-$\sigma$ confidence intervals, and solid lines are the best-fitting solutions. 
    }
    \label{fig_methods:energy}
\end{extfig*}

Finally, it is possible to use IllustrisTNG to control for the effect of large-scale structure in driving the observed quenching directionality by separating satellite galaxies according to when and where they quenched. In particular, as stated in the main text, the observed modulation in the fraction of quiescent satellites could in principle result from processes unrelated to their $z\sim0$ host halo (e.g. pre-processing or quenching as centrals of a different halo). Satellite infall times have been studied for IllustrisTNG satellites \cite{Martina20} and each satellite can be classified into four different groups: star-forming satellites, satellites quenched in their $z\sim0$ host halo, pre-processed satellites (i.e. quenched as satellites in a different halo), and satellites quenched as centrals of a different halo (before being accreted into their $z\sim0$ host). By construction, large-scale structure effects would be responsible for regulating the properties of the two last groups (i.e. pre-processed satellites and those quenched as centrals of a different halo), whereas black hole feedback from the central galaxy and the resulting quenching could only affect either satellites that quenched in their  $z\sim0$ host halo, as well as star-forming satellites (i.e. the first two groups). 

\begin{extfig*} 
    \begin{center}
    \includegraphics[height=6.5cm]{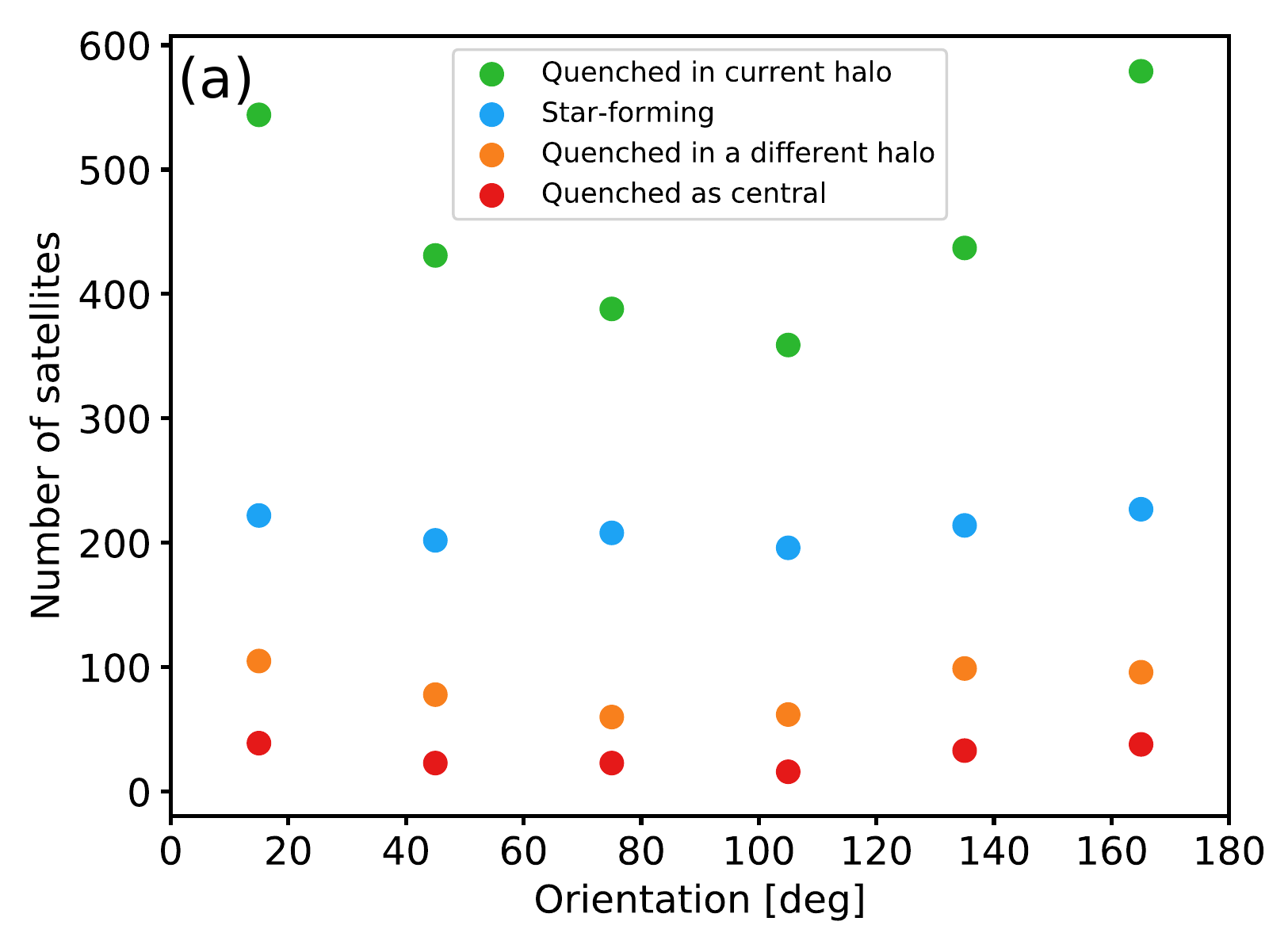}
    \includegraphics[height=6.5cm]{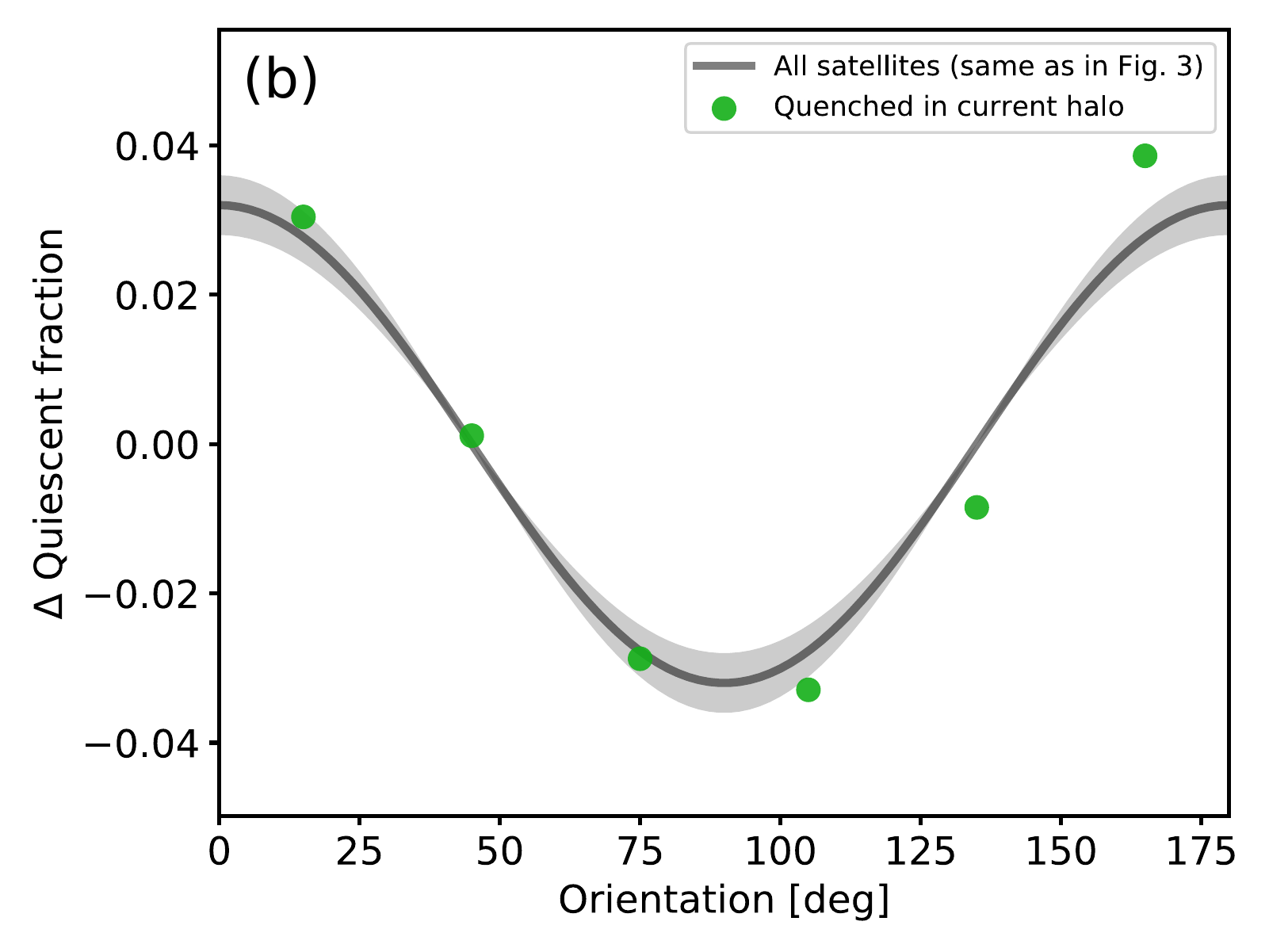}
    \end{center}
    \caption{{\bf Quenching directionality in IllustrisTNG.} Panel (a) shows the number of TNG100 satellites in each orientation bin, depending on whether they are star-forming (blue symbols),  quenched in their $z\sim0$ host halo (green), were pre-processed and quenched in a different halo (orange), or quenched as centrals (red). The last two groups (red and orange symbols) are sensitive to large-scale structure effects, but only correspond to a small fraction of the total satellite population. In panel (b) the fraction of quiescent satellites as a function of orientation is shown but only for those satellites that quenched in their $z\sim0$ host halo (green symbols). The amplitude of this modulation mimics that observed for all IllustrisTNG satellites (grey shaded area and black line).}
    \label{fig_methods:4}
\end{extfig*}

Panel (a) in Extended Data Fig.~\ref{fig_methods:4} shows the number of TNG100 satellites belonging to each group as a function of orientation. It is evident from this figure that,  for the specific galaxy and host mass ranges adopted in this work \cite{Martina20}, large-scale structure-sensitive satellites (red and orange symbols) are outnumbered by both star-forming, and particularly, satellites that quenched in their $z\sim0$ host halo, and thus, large-scale effects could only affect a minority of the satellite population.

Moreover, panel (b) in Extended Data Fig.~\ref{fig_methods:4} shows the modulation in the number of quiescent satellites but only taking into account star-forming satellites and those quenched \textit{within} their $z\sim0$ host halos, compared to the signal observed for the general population of IllustrisTNG satellites. After removing from the analysis those satellites that might be sensitive to large-scale structure effects, the strength of the signal remains the same, following the same trend represented in Fig.~\ref{fig:3} and that observed in the SDSS data. We conclude therefore that the observed modulation in the fraction of IllustrisTNG quiescent satellites is indeed a host halo phenomenon, apparently related to the activity of supermassive black holes in the centers of groups and clusters.

\end{document}